\documentclass[fleqn,usenatbib,useAMS]{mnras}

\usepackage{graphicx}	
\usepackage{amsmath}	
\usepackage{multicol}        
\usepackage{xcolor}

\usepackage[T1]{fontenc}
\usepackage{ae,aecompl}
\usepackage{threeparttable}
\usepackage{orcidlink}

\usepackage{newtxtext,newtxmath}

\title[Ages and abundances of $z\simeq4$ quiescent galaxies]{The \textit{JWST} EXCELS survey: The ages and abundances of $\mathbf{3<z<5}$ massive quiescent galaxies show that downsizing was already in place by $\mathbf{z\simeq4}$}

\author[H-H. Leung et al.]{
Ho-Hin Leung\orcidlink{0000-0003-0486-5178}$^{1}$\thanks{E-mail: hleung2@roe.ac.uk},
Adam C. Carnall\orcidlink{0000-0002-1482-5818}$^{1}$,
Elizabeth Taylor\orcidlink{0000-0001-8728-2984}$^{1}$,
Struan D. Stevenson\orcidlink{0000-0001-5642-752X}$^{1}$,
Aliza G. Beverage\orcidlink{0000-0002-9861-4515}$^{2}$,\and \
Fergus Cullen\orcidlink{0000-0002-3736-476X}$^{1}$,
James S. Dunlop$^{1}$,
Derek J. McLeod\orcidlink{0000-0003-4368-3326}$^{1}$,
Ross J. McLure\orcidlink{0009-0005-9742-2318}$^{1}$,
Ryan Begley\orcidlink{0000-0003-0629-8074}$^{3}$, \and \
Omar Almaini\orcidlink{0000-0001-9328-3991}$^{4}$,
Stella Antonogiannaki$^{1}$,
Karla Z. Arellano-C\'ordova\orcidlink{0000-0002-2644-3518}$^{1}$,
Laia Barrufet\orcidlink{0000-0003-1641-6185}$^{1}$, \and \
Cecilia Bondestam\orcidlink{0009-0008-3775-5112}$^{1}$,
Callum T. Donnan\orcidlink{0000-0002-7622-0208}$^{5}$,
Isaac J. B. Holst\orcidlink{0009-0006-3003-9688}$^{1}$,
Feng-Yuan F. Liu\orcidlink{0000-0003-1386-3676}$^{1}$,
Kate Rowlands\orcidlink{0000-0001-7883-8434}$^{6,7}$, \and \
Ryan L. Sanders\orcidlink{0000-0003-4792-9119}$^{8}$,
Dirk Scholte\orcidlink{0000-0002-6867-1244}$^{1}$,
Maya Skarbinski\orcidlink{0009-0004-0844-0657}$^{6}$,
Thomas M. Stanton\orcidlink{0000-0002-0827-9769}$^{1}$,
and Vivienne Wild\orcidlink{0000-0002-8956-7024}$^{9}$
\\
$^{1}$SUPA\footnote{Scottish Universities Physics Alliance}, Institute for Astronomy, University of Edinburgh, Royal Observatory, Edinburgh EH9 3HJ, UK\\
$^{2}$Department of Astronomy, University of California, Berkeley, CA 94720, USA\\
$^{3}$Armagh Observatory and Planetarium, College Hill, Armagh, BT61 9DG, N. Ireland, UK\\
$^{4}$School of Physics and Astronomy, University of Nottingham, University Park, Nottingham NG7 2RD, UK\\
$^{5}$NSF's National Optical-Infrared Astronomy Research Laboratory, 950 N. Cherry Ave., Tucson, AZ 85719, USA\\
$^{6}$William H. Miller III Department of Physics and Astronomy, Johns Hopkins University, Baltimore, MD 21218, USA\\
$^{7}$AURA for ESA, Space Telescope Science Institute, 3700 San Martin Drive, Baltimore, MD 21218, USA\\
$^{8}$Department of Physics and Astronomy, University of Kentucky, 505 Rose Street, Lexington, KY 40506, USA\\
$^{9}$SUPA, School of Physics \& Astronomy, 
University of St Andrews, 
North Haugh, St Andrews, Fife KY16 9SS, UK
}

\date{Accepted XXX. Received YYY; in original form ZZZ}

\pubyear{{\the\year{}}}

\begin{document}
\label{firstpage}
\pagerange{\pageref{firstpage}--\pageref{lastpage}}
\maketitle

\begin{abstract}
We present deep, medium-resolution $\lambda=1-5\,\mu$m \textit{JWST}/NIRSpec spectroscopy for 14 quiescent galaxies at $3<z<5$ with $\log_{10}(M_*/\mathrm{M_\odot}){\,>\,}10$, obtained as part of the EXCELS survey. We perform a complete re-reduction of these data, including a custom optimal-extraction approach to combat the spectral ``wiggles'' that result from undersampling of the NIRSpec spatial PSF. We constrain the star-formation histories and stellar metallicities of these objects via full-spectral fitting, finding a clear stellar age vs stellar mass correlation, in which more massive galaxies assembled their stellar mass at earlier times. This confirms spectroscopically that the archaeological ``downsizing'' trend was already in place by $z\simeq4$. The slope of our measured relation ($\simeq2$ Gyr per dex in stellar mass) is consistent with literature results at $0 < z < 3$. We do not observe objects with $\log_{10}(M_*/\mathrm{M_\odot})\lesssim10.5$ and ages of more than a few hundred Myr at this epoch, suggesting that recently reported examples of higher-redshift quiescent galaxies at these masses are likely to soon rejuvenate. We measure relatively high stellar metallicities for the majority of our sample, consistent with similar objects at $0 < z < 3$. Finally, we explore evidence for $\alpha$-enhancement in six older and more luminous galaxies within our sample, finding considerable disagreements in the chemical abundances measured using different stellar population models, different fitted rest-frame wavelength ranges, star-formation history models and fitting codes. We therefore conclude that inferring detailed stellar chemical abundances for the earliest quiescent galaxies remains challenging, and higher signal-to-noise spectra are required (observed frame SNR per resolution element $\simeq100$ for $R\simeq1000$).

\end{abstract}


\section{Introduction}\label{sec:intro}

The most massive galaxies in the Universe provide a unique constraint on the process of galaxy formation, as they represent a limiting case for the underlying physics. In particular, extending the study of the most massive galaxies to progressively higher redshifts has repeatedly revolutionised our perspective on galaxy formation across the last several decades (e.g., \citealt{Dunlop1996, Cimatti2004, Daddi2005}).

As early as the mid-to-late 2000s, indications began to emerge of galaxies at $z>3$ that had already reached high stellar masses and subsequently shut down, or quenched, their star-formation activity (e.g., \citealt{Caputi2004, Fontana2009}). Improvements in the available datasets led to large and robust photometric samples beginning to emerge in the mid-to-late 2010s (e.g., \citealt{Straatman2014, Straatman2016, Merlin2018, Merlin2019, Carnall2020}), with spectroscopic confirmation following shortly thereafter (e.g., \citealt{Glazebrook2017, Schreiber2018}).

The advent of the \textit{James Webb Space Telescope} (\textit{JWST}) in 2022 quickly produced an explosion in the number of photometric candidate massive quiescent galaxies at $z>3$ (e.g., \citealt{Carnall2023b, Valentino2023}), as well as in the quality and quantity of spectroscopic follow-up observations available (e.g., \citealt{Glazebrook2023,Nanayakkara2024,Nanayakkara2025}), including the first spectroscopic confirmations significantly beyond $z=4$ (e.g., \citealt{Carnall2023c, UrbanoStawinski2024, deGraaff2025}).

Extensive theoretical work has been undertaken to try to explain both the formation of so much stellar mass within such a short space of time (e.g., \citealt{Dekel2023, Dekel2025, Silk2024, Shen2024}), as well as the rapid quenching of star formation in the gas-rich high-redshift Universe (e.g., \citealt{Hartley2023, De_Lucia2024, Lagos2025, Kimmig2025, Chandro-Gomez2025}). Significant interest has also developed into the fates of these extreme objects, for example whether they go on to form the cores of the most massive galaxies in the local Universe (e.g., \citealt{Baggen2023, Beverage2024, Rennehan2024, Remus2025, Baker2025a}), fuelled by early hints that such objects tend to exist within overdense environments (e.g., \citealt{Jin2024,Alberts2024,Jespersen2025,McConachie2025}).

However, such work has so far been guided only by very limited observational evidence. Robust, \textit{JWST}-derived number densities for $z>3$ massive quiescent galaxies have only been available from a handful of small area studies ($\lesssim150$ sq. arcmin, e.g., \citealt{Carnall2023b, Valentino2023, Long2023, Alberts2024, Russell2025, Baker2025b}), with the first larger-area studies only very recently becoming available ($\gtrsim300$ sq. arcmin, e.g., \citealt{Baker2025c, Merlin2025, Stevenson2026}). More-detailed physical properties derived from spectroscopy are so far only available for individual bright and potentially unrepresentative objects (e.g., \citealt{Carnall2023c, Setton2024, Wu2025, deGraaff2025}).

Furthermore, much of the \textit{JWST} spectroscopic data available so far at ${z>3}$ is taken with the \textit{JWST}/NIRSpec PRISM mode (${R=\lambda/\Delta\lambda\simeq30-300}$), far below the ${R\simeq1000}$ required to robustly derive physical properties such as stellar ages and metallicities, using detailed study of individual spectral features (e.g., \citealt{Ocvirk2006, Pacifici2012}). To probe in detail the mechanisms that result in the growth and quenching of early massive galaxies at ${z>3}$, larger and more-representative galaxy samples with high signal-to-noise ratio (SNR) medium-resolution continuum spectroscopy are required.

A broad range of \textit{JWST} medium-resolution high-SNR continuum spectroscopy already exists for massive ($\log_{10}(M_*/
\mathrm{M_\odot})>10$) quiescent galaxies at $z\simeq1-3$, and is being widely exploited, demonstrating the value of such data (e.g., \citealt{Belli2023, Kriek2024, Park2024, Davies2024, Slob2024, Slob2025, Beverage2025, Bugiani2025, Ito2025b, Skarbinski2026, Hamadouche2026}). Such works build upon ground-based efforts over many years to characterise the stellar populations in massive galaxies outside of the local Universe via deep optical-NIR continuum spectroscopy (e.g., \citealt{vanderwel2016, McLure2018b, Pentericci2018, Wild2020, Beverage2024}).

One of the most fundamental results of this work is the archaeological downsizing trend. At fixed redshift, more-massive galaxies are found to have formed their stellar populations earlier in cosmic history than their less-massive counterparts (e.g., \citealt{Gallazzi2005, Gallazzi2014, Belli2019, Carnall2019b, Beverage2021,Hamadouche2023,Slob2024,Merlin2025}).
Downsizing is thought to arise primarily due to the baryon-to-star conversion efficiency in galaxy halos being a strong function of halo mass \citep[e.g.,][]{Moster2018}, with halos of masses $M_\mathrm{h}\sim10^{12}\,\mathrm{M_\odot}$ being the most efficient. This means that the most massive halos at any given epoch typically passed through this highly efficient mass regime earlier than less-massive halos, forming the bulk of their stars during this time. There is also evidence that the peak conversion efficiency evolves modestly towards lower masses at later times, exacerbating this effect \citep{Behroozi2010}.

An alternative, more-empirical perspective on this is provided by \cite{Abramson2016}, who argue that downsizing is a natural consequence of the redshift evolution of the star-formation main sequence \citep[SFMS; e.g.,][]{Speagle2014, Popesso2023}. Higher star-formation rates (SFRs) at all stellar masses at earlier times and the less-than-unity slope of the SFMS ($\rm{d}\log \mathrm{SFR}/\rm{d}\log M_*<1$) mean that early-forming galaxies experienced rapid growth in stellar mass, quickly reaching relatively low star-formation efficiencies (i.e., quenching) at the high-mass end of the SFMS. For galaxies that formed later, lower SFRs at all stellar masses lead to slower mass assembly and thus more-extended star-formation histories (SFHs).

Dry mergers after quenching have also been suggested as a contributing factor to downsizing \citep{Cattaneo2008}. Galaxies that assembled and quenched earlier are often central galaxies in groups and clusters, and so have many opportunities to further grow in mass via dry mergers. Other ideas centred on quenching mechanisms that could become less effective at high-redshift have also been proposed to contribute to downsizing, such as active galactic nucleus (AGN) feedback (e.g., \citealt{Scannapieco2005, Croton2006}), virial-shock heating (e.g., \citealt{Dekel2006, Rodriguez-Puebla2017}) and environmental effects (e.g., \citealt{Taylor2023}).

Further key insights into the formation and quenching of massive quiescent galaxies can be gained by the study of their stellar metallicities. Foundational work in the local Universe established the existence of a stellar metallicity vs stellar mass relationship (e.g., \citealt{Gallazzi2005, Panter2008}), which flattens at the highest masses. The fact that local star-forming galaxies display lower stellar metallicities than their quiescent counterparts at fixed stellar mass has been interpreted as suggesting that an extended process of gas exhaustion over several billion years is the primary mechanism that quenched low-redshift quiescent galaxies (e.g., \citealt{Peng2015,Trussler2020}; see also \citealt{Vaughan2022,Baker2024,Looser2024}). However the many free parameters and degeneracies involved in even the simplest one-zone analytic chemical evolution models make the interpretation of such results extremely challenging. For example, recent results from \cite{Leung2024} demonstrate that an increase in stellar metallicity can also occur as a result of an intense, short-timescale starburst event immediately prior to quenching.

More recently, the analysis of quiescent galaxy stellar metallicities has been extended to higher redshift (e.g., \citealt{Gallazzi2014, Kriek2016, Kriek2019, Carnall2019b, Carnall2022b, Jafariyazani2020, Beverage2024, Cheng2024, Hamadouche2026}). The long timescales of many billions of years involved in gas exhaustion models immediately suggests an alternative quenching mechanism must dominate at the cosmic noon epoch and beyond ($z\gtrsim1$), with many studies focusing on the 
possibility of rapid quenching via gas expulsion driven by strong AGN feedback (e.g., \citealt{Beverage2021}). Evidence for mild evolution in the quiescent stellar mass vs stellar metallicity relation towards higher redshifts is mixed, however strong evolution is now effectively ruled out, at least as far as $z\simeq3$ \citep{Lonoce2015,Onodera2015,Estrada-Carpenter2019,Beverage2025,Hamadouche2026}.

Extending this substantial body of work beyond $z=3$ is the primary motivation for the \textit{JWST} Cycle 2 Early eXtragalactic Continuum and Emission-Line Science (EXCELS) Survey \citep[][Programme ID: 3545, PI: Carnall; Co-PI: Cullen]{Carnall2024}. EXCELS targeted a representative sample of 14 candidate massive quiescent galaxies at $3 < z < 5$ in the PRIMER UDS field, obtaining ultra-deep, medium-resolution continuum spectroscopy from $\lambda=1-5\,\mu$m for all these objects. In this work, we present this representative spectroscopic sample and perform full spectrophotometric fitting, in combination with the available photometry, to derive detailed physical properties for these objects. In particular, we focus on their stellar ages and metallicities. In companion papers, we investigate the processes giving rise to line emission in these objects \citep{Stevenson2026}, as well as evidence for neutral outflowing gas via absorption line analysis \citep{Taylor2026}.

This paper is structured as follows. In Section \ref{sec:method} we introduce our photometric and spectroscopic data and reduction methodology. We also present our fiducial full-spectral-fitting methodology using the \textsc{Bagpipes} code and \cite{Bruzual2003} stellar population models, by which our main results are derived. These results are presented in Section \ref{sec:results}. In Section \ref{sec:alpha} we investigate alternative full-spectral-fitting approaches using stellar population model libraries that allow for enhanced $\alpha$-element abundances, instead of the standard approach of assuming scaled-solar abundances for all elements. We present our conclusions in Section \ref{sec:conclusion}.

All magnitudes are quoted in the AB system \citep{Oke1983}. For cosmological calculations, we adopt $\mathrm{\Omega_m} = 0.3$, $\mathrm{\Omega_\Lambda} = 0.7$, and $\mathrm{H_0 = 70 \,km\, s^{-1}\,Mpc^{-1}}$. We assume a \citet{Kroupa2001} stellar initial mass function (IMF), and take solar metallicity $Z_\odot=0.0142$ \citep{Asplund2009}. We re-scale all metallicity measurements quoted from the literature to this solar metallicity value for direct comparison. Due to the rapidly evolving nature of these early galaxies, we average SFRs over a 10 Myr period throughout.

\section{Data Reduction and Fitting Methodology}\label{sec:method}

\subsection{PRIMER photometric data}\label{sec:photometry} 

The primary photometric dataset supporting this work is the large public \textit{JWST} Cycle 1 programme Public Release IMaging for Extragalactic Research (PRIMER; Dunlop et al. in prep.; \citealt{PRIMER_proposal}), which provides contiguous 8-band NIRCam imaging (F090W, F115W, F150W, F200W, F277W, F356W, F410M and F444W) plus 2-band MIRI imaging (F770W and F1800W) in the UDS and COSMOS fields.

The photometric data used in this work comes from the PRIMER UDS catalogue described in \cite{Begley2025}. This is an updated version of the photometric catalogue used to select the EXCELS sample, as described in \cite{Carnall2024}. In this section we present a brief overview of the \cite{Begley2025} photometric catalogue. 

\cite{Begley2025} reduce the available NIRCam data using the PENCIL software (PRIMER Enhanced NIRCam Image-processing Library; Magee et al. in prep.), which builds on the standard \textit{JWST} data reduction pipeline (v1.10.8; pmap$>$1118). PENCIL includes custom additional routines for background subtraction, 1/f noise correction and snowball/wisp removal.

The resulting NIRCam mosaics are then PSF-homogenised to the F444W band using kernels constructed by stacking bright, unsaturated stars in the PRIMER field. The PRIMER data is also supplemented with deep optical \textit{HST}/ACS F435W, F606W and F814W imaging \citep{Grogin2011, koekemoer2011}. The EXCELS sample was selected only within the area that has coverage in all three \textit{HST} bands, meaning we have full $\lambda=0.4-5\,\mu$m coverage for all our objects. We use the \textit{HST} mosaics produced as part of the Hubble Legacy Field project \citep{Illingworth2016, Whitaker2019}. These \textit{HST} data are also similarly PSF-homogenised to the PRIMER F444W-band imaging.

The multi-wavelength catalogue is then constructed using Source Extractor \citep{Bertin1996} in dual image mode, with the unconvolved F356W mosaic used as the detection band in all cases. Fluxes are extracted in $0.5^{\prime\prime}-$diameter apertures, and these aperture fluxes are corrected to total using the FLUX\_AUTO output from Source Extractor \citep{Kron1980}. We then apply a further 10 per cent correction following \cite{McLeod2024} to account for flux not included within the Kron aperture.

The 14 EXCELS quiescent galaxies at $z>3$ selected by \cite{Carnall2024} are all recovered from this new catalogue via a close positional match. Their celestial coordinates and spectroscopic redshift are listed in Table 1 of \cite{Carnall2024}. All 14 objects have coverage in all 11 \textit{HST}+NIRCam photometric bands described above. Additionally, as demonstrated in \cite{Stevenson2026}, all 14 are still selected from the new \cite{Begley2025} catalogue as $z>3$ massive quiescent galaxy candidates via a similar process to that employed by \cite{Carnall2024}.

\cite{Stevenson2026} also consider mass-completeness limits for their sample of photometrically selected $z>3$ massive quiescent galaxy candidates (see their fig. 1), concluding that their F356W < 26 sample is more than 99 per cent mass complete at log$_{10}(M_*/\mathrm{M_\odot}) > 10$. The EXCELS sample is selected requiring F356W < 25, meaning that, following \cite{Stevenson2026}, it is mass complete at log$_{10}(M_*/\mathrm{M_\odot}) > 10.4$. This means that our sample is fully representative of massive quiescent galaxies across this redshift and mass range. As discussed in Section \ref{sec:results}, the stellar masses we derive for all 12 of the 14 EXCELS objects for which we are able to obtain robust spectral fits are above this threshold.

\subsection{EXCELS spectroscopic data}
\label{sec:excels_reduction}

The EXCELS dataset is described in full in \cite{Carnall2024}; here we provide a brief summary. The survey consists of 4 NIRSpec pointings within the PRIMER UDS imaging area, which were each observed with all 3 medium-resolution gratings using different mask configurations in order to maximise the number of objects for which rest-frame optical coverage could be obtained. The exposure times obtained were $\simeq4$ hours in G140M, $\simeq5.5$ hours in G235M and $\simeq4$ hours in G395M. As the top-priority targets for the survey, the 14 objects that are the focus of this work were observed in all three gratings, providing continuous wavelength coverage from $1-5\,\mu$m (excepting objects with traces that cross the NIRSpec detector gap, which have short interruptions in coverage).

\subsubsection{Initial spectroscopic reduction with the NIRSpec pipeline}\label{sec:pipeline}

We perform a new, customised reduction of the 14 EXCELS spectra for massive quiescent galaxies at $z>3$ using a modified version of the standard \textit{JWST} reduction pipeline\footnote{\url{https://github.com/spacetelescope/jwst}} \citep[][version 1.19.1]{JWST_pipeline}. We use the CRDS\_CTX = \texttt{jwst\_1413.pmap} version of the \textit{JWST} Calibration Reference Data System (CRDS) files.

First, we obtain the level 1 ``uncal'' products from the Mikulski Archive for Space Telescopes (MAST). These products are processed by the level 1 pipeline (\texttt{calwebb\_detector1}) using default configurations except we turn on the \texttt{clean\_flicker\_noise} step to remove the effects of 1/f noise from the exposures. For this reduction step, we set \texttt{fit\_method} to \texttt{median}, \texttt{mask\_science\_region} to \texttt{True}, \texttt{background\_method} to \texttt{None} and \texttt{n\_sigma} to 2. These choices were arrived at after extensive experimentation to determine which configuration produces the cleanest results in unilluminated areas of the detector.

We next run the resulting level 2a ``rate'' products through the level~2 pipeline (\texttt{calwebb\_spec2}) using the default configuration options. Background subtraction is performed on each of the three nod positions by subtracting the sigma-clipped average of the other two nod positions (using the default 3 sigma threshold). For all observations of PRIMER-EXCELS-52467 and for two nods in the G140M grating observations of PRIMER-EXCELS-112990 only one background nod was used, in order to avoid contamination in the other background shutter.

We then mask any pixels in the resulting level 2b ``cal'' files flagged with any among a customised list of data quality bit-masks\footnote{Bits: 0, 1, 3, 4, 6, 7, 10, 11, 12, 13, 16, 17, 18, 19, 20, 21, 24, 26, 27, 29 and 30 (see \url{https://jwst-pipeline.readthedocs.io/en/latest/jwst/references_general/references_general.html\#data-quality-flags})}. We also perform additional manual masking of bad pixels, snowballs not caught by the automatic detection step in the level 1 pipeline, bleeding from hot pixels, and contamination from overlapping open shutters or bright stars. Following this, we pass all ``cal'' products into the level 3 pipeline (\texttt{calwebb\_spec3}) using the default configuration options.

\begin{figure}
    \centering
    \includegraphics[width=\columnwidth]{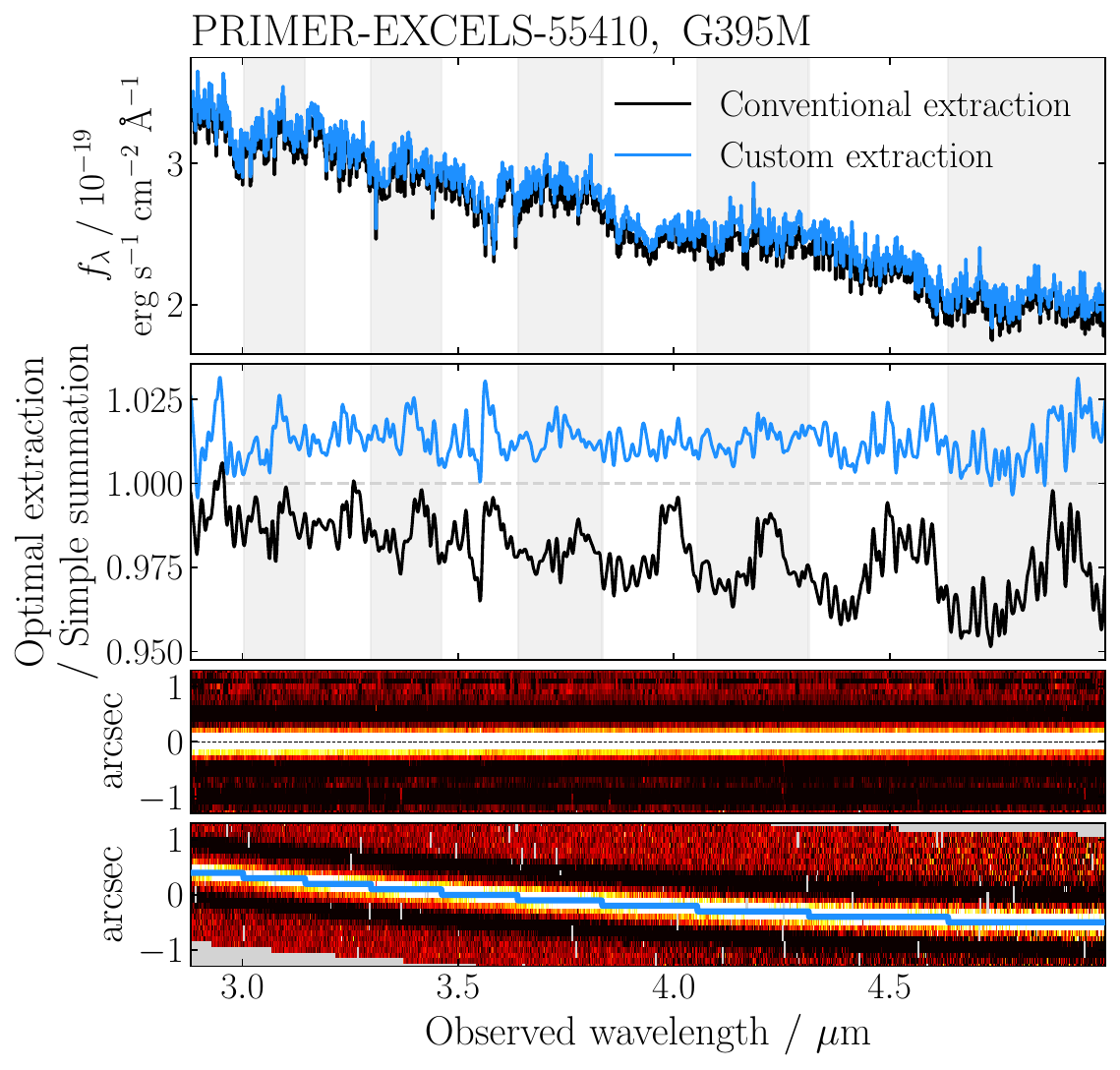}
    \caption{A comparison of different optimal extraction methods applied to our NIRSpec data (see Section \ref{sec:wiggles}). Our custom wavelength-varying 1D optimal extraction is shown in blue; the more-conventional fixed-kernel method is shown in black. The top panel shows the 1D extracted spectrum from the G395M grating for PRIMER-EXCELS-55410 (ZF-UDS-7329), extracted using the two methods. The second panel shows the ratio between the extracted spectrum and the spectrum obtained from simply summing the central 5 rows of the 2D spectrum (roughly the projected size of one shutter). The third panel shows the level 3 pipeline output 2D spectrum (after \texttt{drizzle} resampling), and the bottom panel shows the level 2 2D spectrum for one of the three nod positions (before \texttt{drizzle} resampling). The pixel containing the object centroid is marked with a blue line in the bottom panel. In the top two panels we also colour with alternating grey and white bands wavelength ranges for which the centroid of the trace falls on the same pixel row in the bottom panel. Periodic fluctuations at a $\simeq5$ per cent level can be seen in the black spectrum, with the pattern repeating every time the centroid shifts to the next pixel down in the bottom panel. Our custom optimal extraction method mitigates these fluctuations by using a wavelength-varying extraction kernel.}
    \label{fig:wiggles}
\end{figure}

\begin{figure*}
    \centering
    \includegraphics[width=0.98\textwidth]{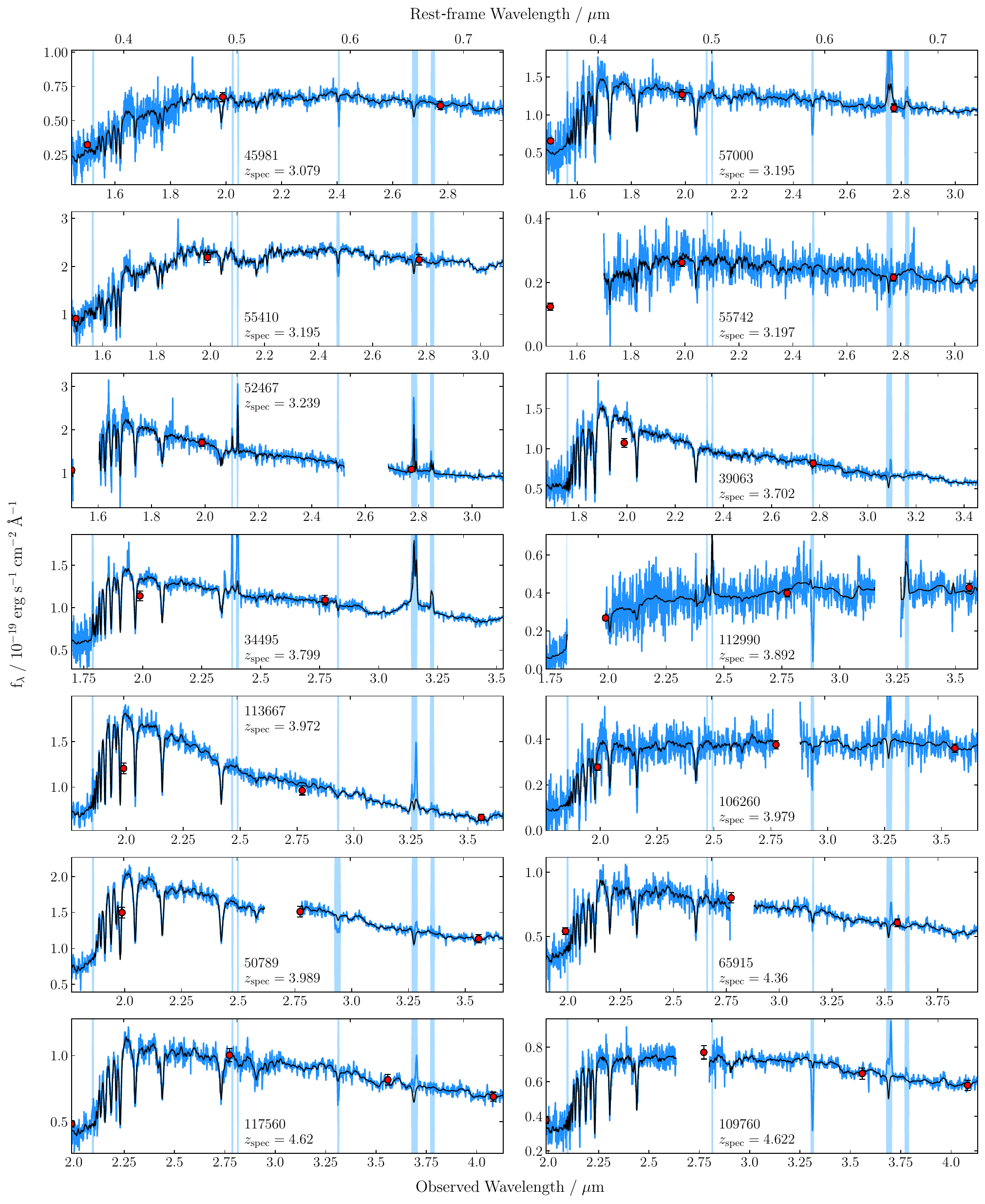}
    \caption{The spectral energy distributions (SEDs) and spectra for the 14 massive $z>3$ quiescent galaxies observed as part of EXCELS. The calibrated EXCELS NIRSpec observations within the rest-frame wavelength range $3540-7350$ {\AA} used in our primary fitting methodology are shown in blue. PRIMER NIRCam photometry within this wavelength range are shown as red dots. The posterior median model spectra fitted using \textsc{Bagpipes} are shown in black. Vertical blue bars mark the regions masked during fitting.}
    \label{fig:all_spec}
\end{figure*}

\begin{table*}
\centering
\caption{Model priors used when performing Bayesian fits of the EXCELS spectra jointly with \textit{HST}+\textit{JWST} photometry (see Section \ref{sec:fitting} for a full description). Some parameters have prior shape $\log_{10}$ uniform, which indicates a flat prior in logarithmic space $\log(X) \sim U(\log(\mathrm{min}), \log(\mathrm{max}))$. Note that $\sigma_\mathrm{disp}$ is not the intrinsic velocity dispersion of the galaxy, as it does not account for the finite resolution of the spectral templates or observational data.}
\begin{threeparttable}
    \begin{tabular}{lllll}
        \hline
        Type & Parameter & Form & Min & Max  \\
        \hline
        SFH & $\log_{10}(M_*/\mathrm{M_\odot})$ & Uniform & 0 & 13 \\
         & Stellar metallicity $Z_*/Z_\odot$ & Uniform & 0.007 & 3.52\tnote{\textdagger} \\
         & Double-power-law falling slope $\alpha$ & $\log_{10}$ Uniform & 0.1 & 1000 \\
         & Double-power-law rising slope $\beta$ & $\log_{10}$ Uniform & 0.1 & 1000 \\
         & Double-power-law turnover time $\tau$ / Gyr & Uniform & 0.1 & $t_\mathrm{obs}$ \\
         \hline
        Dust & $A_V$ / mag & Uniform & 0 & 4 \\
         & Deviation from Calzetti slope $\delta$ & Gaussian ($\mu=0$, $\sigma=0.1$) & -0.3 & 0.3 \\
         & Strength of 2175{\AA} bump $B$ & Uniform & 0 & 5 \\
         & Birth cloud factor $\eta$ & Fixed = 2 & - & - \\
         \hline
        AGN\tnote{*} & Continuum flux at 5100{\AA} $f_{5100}$ / $\mathrm{erg\, s^{-1}\, cm^{-2}\, {\text\AA}^{-1}}$ & Uniform & 0 & $10^{-19}$ \\
         & Spectral index at $\lambda_\mathrm{rest}<5100\text{\AA}$ $\alpha_{\lambda,<5000}$ & Gaussian ($\mu=-1.5$, $\sigma=0.5$) & -2 & 2 \\
         & Spectral index at $\lambda_\mathrm{rest}>5100\text{\AA}$ $\alpha_{\lambda,>5000}$ & Gaussian ($\mu=0.5$, $\sigma=0.5$) & -2 & 2 \\
         & Broad H$\alpha$ normalisation $f_{\mathrm{H}\alpha, \mathrm{broad}}$ / $\mathrm{erg\, s^{-1}\, cm^{-2}}$ & Uniform & 0 & $2.5\times10^{-17}$ \\
         & Broad line velocity dispersion $\sigma_\mathrm{AGN}$ / km/s & $\log_{10}$ Uniform & 1000 & 7000 \\
         \hline
        GP noise & Uncorrelated amplitude (white noise scaling) $s$ & $\log_{10}$ Uniform & 0.1 & 10 \\
         & Correlated amplitude $\sigma$ & $\log_{10}$ Uniform & $10^{-4}$ & $10^{-1}$ \\
         & Period/length scale $\rho$ & $\log_{10}$ Uniform & 0.04 & 4.0 \\
         & Dampening quality factor $Q$ & Fixed = 0.49 & - & - \\
         \hline
        Miscellaneous & Redshift & Gaussian ($\mu=z_\mathrm{spec}$, $\sigma=0.01$) & $z_\mathrm{spec}-0.05$ & $z_\mathrm{spec}+0.05$ \\
         & Stellar velocity dispersion $\sigma_{\mathrm{disp}}$ / km/s & $\log_{10}$ Uniform & 50 & 500 \\ \hline
    \end{tabular}
    \begin{tablenotes}
        \item [\textdagger] Prior limits for stellar metallicity are 0.005-2.5 $\mathrm{Z_\odot}$ for $\mathrm{Z_\odot}=0.02$ as assumed in the \cite{Bruzual2003} stellar population models. For our solar metallicity value at $\mathrm{Z_\odot}=0.0142$, this converts to 0.007-3.52 $\mathrm{Z_\odot}$.
        \item [*] Component not included when fitting galaxies 45981, 50789, 55410, 55742, 65915 and 117560.
    \end{tablenotes}
\end{threeparttable}
\label{tab:priors}
\end{table*}

\subsubsection{Optimal 1D spectral extraction}\label{sec:wiggles}

We extract 1D spectra from the level 3 output ``s2d'' files via optimal extraction \citep{Horne1986}. When initially performing optimal extraction using the same fixed kernel at all wavelengths (constructed via Gaussian fits to the wavelength-collapsed 2D spectra), we noticed $\simeq5$ per cent systematic fluctuations (``wiggles'') in the continua of the extracted spectra \citep[e.g.,][]{Perna2023,Dumont2025}. This effect is demonstrated in Fig. \ref{fig:wiggles}, where our initial fixed-kernel optimal extraction run is compared with a simple sum over the central 5 rows of the rectified 2D spectrum (``s2d'' output) for an example object. The top panel shows the fixed-kernel extraction in black, and the ratio of the fixed-kernel extraction to the simple sum extraction is shown in black in the middle panel. The optimal extraction is expected to produce a very similar spectrum to the simple sum, with higher SNR, however periodic wiggles are clearly visible.

The bottom panels of Fig. \ref{fig:wiggles} show the rectified ``s2d'' (above) and unrectified ``cal'' (below) 2D spectra for this object. The former is a stack of all three nod positions, the latter is for a single example nod position. It can be seen that the wiggles in the black line in the middle panel have periods in wavelength that exactly match the pattern of the object's centroid shifting from one row to the next in the unrectified (bottom) 2D frame, shown by the blue horizontal line.

This pattern arises due to the \texttt{resample\_spec} step in the level 3 pipeline, which uses the \texttt{drizzle} algorithm to resample the ``cal'' products from the three nod positions into a single rectified product \citep{Fruchter2002}. The algorithm by default assumes that flux is uniformly distributed within each pixel, since the true distribution of flux at higher spatial resolution is not known. The pixel scale for NIRSpec is $0.1^{\prime\prime}$, meaning the spatial point spread function is significantly undersampled, and so the assumption of uniformly distributed flux within pixels is not valid for compact sources such as ours. This leads to different results depending on where the object centroid falls relative to the spatial pixels. For example, the object trace is wider in the spatial direction when the centroid falls on the boundary between two pixels, and narrower when the centroid falls in the centre of a pixel. This means that our assumption of a single fixed-width kernel at all wavelengths is a poor representation of the data, leading to the $\simeq5$ per cent systematic fluctuations seen in Fig. \ref{fig:wiggles}. 

We therefore instead perform a custom wavelength-varying 1D optimal extraction of our 2D spectra. We begin by fitting a Gaussian to the wavelength-collapsed 2D spectrum, as previously, to provide an initial estimate of the centroid position and $\sigma$ width. Next, we isolate only the rows in the 2D spectra that are within $\pm3\sigma$ of the fitted centroid. We then loop over all wavelength bins, summing in the wavelength direction within a boxcar of $\pm25$ pixels ($\simeq300$, 500 and 900 {\AA} for G140M, G235M and G395M, respectively). Within each iteration, we then calculate the total SNR of the wavelength-collapsed spectrum slice.
If $\mathrm{SNR}\geq5$, we flux normalise the wavelength-collapsed spectrum slice and use this directly as the optimal extraction weights in this wavelength bin. If $\mathrm{SNR}<5$, we instead use the initial Gaussian fit over all wavelengths for our optimal extraction weights at this wavelength. The code for this extraction method have been made available online\footnote{\url{https://github.com/HinLeung622/rolling_1D_extraction}}.

An example result from this custom extraction method is shown in blue in Fig. \ref{fig:wiggles}, and is compared against the conventional method described above, which is shown in black. It can be seen that the fluctuations in sync with the shifting of the trace's centroid have been removed. Thus, the amplitude of the systematic fluctuations with respect to the simple sum over 5 pixel rows has been reduced to $\simeq1$ per cent.

\subsubsection{Combination of separate gratings and correcting slit loss}

To combine 1D spectra from the three separate gratings for each object, we first calculate the mean flux in the overlapping wavelength regions, then scale the G140M and G395M spectra to the normalisation of the G235M data. In the absence of any detector gaps in the overlapping regions, we degrade the resolution of the higher-resolution (shorter-wavelength) grating in both overlapping regions to match the lower-resolution grating using \textsc{SpectRes} \citep{Carnall2017}. If a detector gap is present in one of the gratings within the overlapping region, we instead resample the grating with the gap to match the resolution of the grating without the gap. Lastly, all overlapping regions are combined through taking a simple mean of the pixel values from the two gratings. Due to their very noisy G140M spectra, we only combine the G235M and G395M spectra for three galaxies (55742, 106260 and 112990).

To account for slit losses and potentially imperfect spectrophotometric calibration, we then scale the joined spectrum for each object to match the best fit model from an initial fit to only the \textit{HST}+\textit{JWST} photometry described in Section \ref{sec:photometry}. The fit is performed using the Bayesian spectral fitting code \textsc{Bagpipes} \citep{Carnall2018,Carnall2019b}, with an almost identical model configuration and priors to those described in Section \ref{sec:fitting}, except we remove the spectroscopy-specific Gaussian process and velocity dispersion components, and do not include AGN contributions to limit complexity. We also fix the redshifts to the values measured manually by the EXCELS team, as described in \cite{Carnall2024}. 

To scale our joined spectra to these best-fit models, we first create a temporary degraded EXCELS spectrum from the joined gratings by binning in groups of 5 pixels. The best-fit model spectrum is then resampled onto the same coarse wavelength grid using \textsc{SpectRes}. We next calculate the ratio between the resampled model spectrum and a rolling median of the degraded EXCELS spectrum, calculated using a window that spans 101 bins ($\sim3000-9000$ {\AA}). These smoothing steps are performed to prevent the recalibration from altering the spectral shape on short wavelength scales, which risks altering individual emission/absorption line equivalent widths or spectral break strengths. 

Finally, we fit a 15\textsuperscript{th} order Chebyshev polynomial to these calculated ratios as a function of wavelength \citep[e.g.,][]{Cappellari2017}, and recalibrate our joined EXCELS spectra according to this polynomial. Our final joined and calibrated spectra for the 14 EXCELS quiescent galaxies at $3<z<5$ are shown in Fig. \ref{fig:all_spec}.

\begin{figure*}
    \centering
    \includegraphics[width=\textwidth]{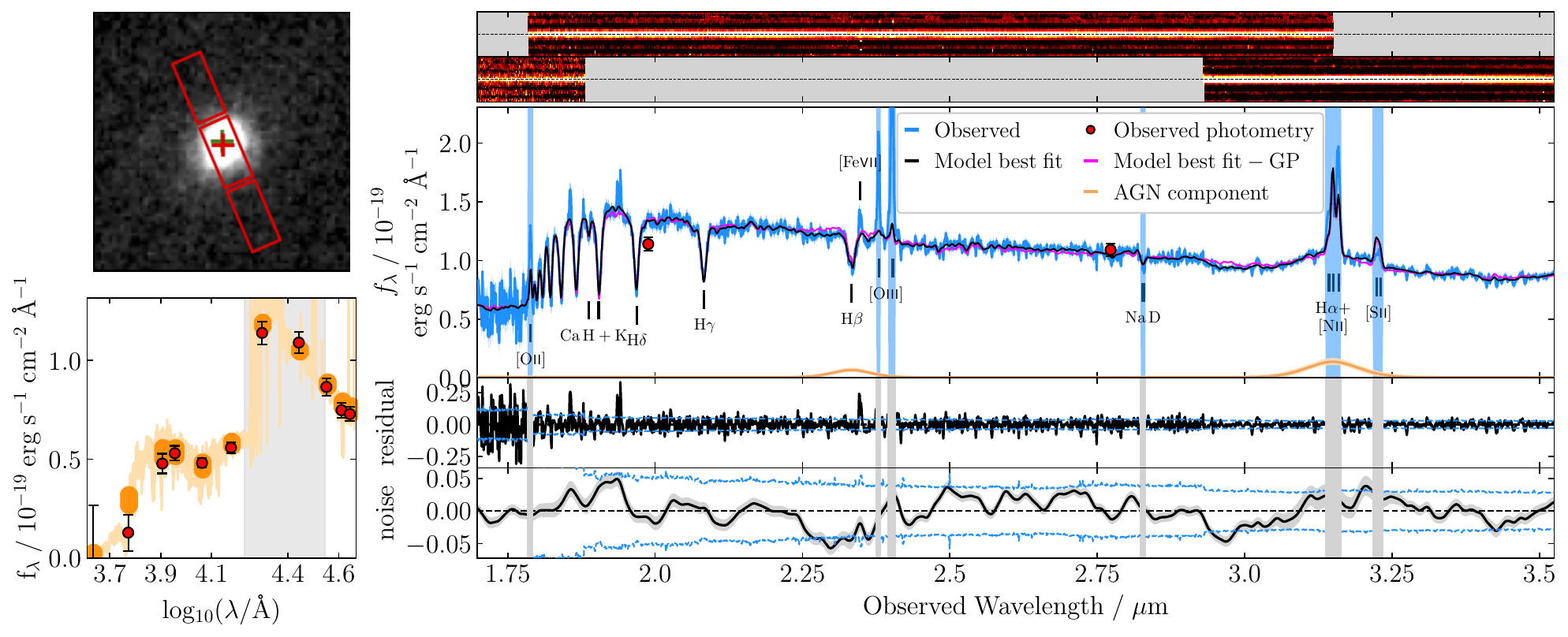}
    \caption{Detailed \textsc{Bagpipes} full-spectral fitting of PRIMER-EXCELS-34495, shown as an example of the process described in Section \ref{sec:fitting}. \textbf{Top left}: A PRIMER F277W cutout image of the galaxy. The overlaid rectangles mark the MSA slit positions for the first of the three G235M nod positions, while the cross marks the extraction centroid (see Section \ref{sec:wiggles}) \textbf{Top right}: Observed 2D spectra for the three medium resolution gratings. \textbf{Central right}: Our 1D spectroscopic extraction within the rest-frame wavelength range $3540{-}7350$ {\AA} (blue) and \textit{HST}+\textit{JWST} observed photometry (red points). Plotted on top are the fitted posterior median model spectrum (black line), the physical model spectrum (magenta line: posterior median model with the posterior median GP component subtracted) and the fitted AGN component (orange line). \textbf{Lower right}: Residuals (black) between our fitted model and data, along with the input observational uncertainties (dotted blue lines) scaled by our $s$ parameter. \textbf{Bottom right}: The fitted additive GP noise component in black, along with the same scaled input observational uncertainties. The y-axes of both the residual and noise panels have the same units as the central panel, but an expanded scale. In the central, residual and noise panels, the vertical shaded bars mark the regions masked during fitting. \textbf{Bottom left}: The full observed multi-band photometry (red points), described in Section \ref{sec:photometry}, and the corresponding model posteriors (orange patches). The orange curve shows the posterior median model spectrum and the shaded region marks the rest-frame wavelength range from $3540{-}7350$ {\AA}.}
    \label{fig:spec_fit_example}
\end{figure*}

\subsection{Bayesian spectrophotometric fitting} \label{sec:fitting}

To measure galaxy properties, we then perform Bayesian full spectral fitting of the spectra for our 14 EXCELS massive quiescent galaxies at $3<z<5$, reduced and calibrated as described in Section \ref{sec:excels_reduction}, in combination with the \textit{HST}+\textit{JWST} photometry described in Section \ref{sec:photometry}. Fitting is performed using \textsc{Bagpipes} (version 1.3.5), with a similar model configuration and priors to those adopted in \cite{Carnall2024}. In Table \ref{tab:priors}, we list all fitted model parameters and their priors. All input spectra are first truncated to rest-frame wavelengths from $3540-7350$ {\AA}  before being passed into \textsc{Bagpipes}, which is the wavelength range spanned by the empirical and high-spectral-resolution MILES library in the \cite{Bruzual2003} models. We additionally mask the [O\,\textsc{ii}], [O\,\textsc{iii}], [N\,\textsc{ii}], [S\,\textsc{ii}] and H$\alpha$ emission lines (but not H$\beta$), as well as the Na D absorption line, due to potential non-stellar contributions.

\subsubsection{Bagpipes full spectral fitting approach}
Within \textsc{Bagpipes}, we employ the \cite{Bruzual2003} stellar population synthesis models (2016 version, described in \citealt{Chevallard2016}), which incorporates the high-resolution rest-frame optical empirical stellar spectral templates from the MILES library \citep{Sanchez-Blazquez2006,Falcon-Barroso2011}. We assume the initial mass function of \cite{Kroupa2001}, and a double-power-law SFH model \citep[e.g.,][]{Carnall2019a}. We assume a uniform and time-invariant stellar metallicity for each galaxy, which is allowed to vary with a uniform prior.

We model nebular emission using the \textsc{Cloudy} photoionization code \citep{Ferland2017}, with an approach based on that of \cite{Byler2017}. The nebular metallicity is fixed to the stellar value, and the ionization parameter is held fixed at log$_{10}(U)=-3$.

We model dust attenuation using the variable-slope model from \cite{Salim2018}, which is based on a power-law perturbation of the \cite{Calzetti2000} dust law. Any stars younger than $10\,$Myr are assumed to be more attenuated than older stars by a factor $\eta=2$, as they are assumed to still be surrounded by their birth clouds. We model intergalactic medium attenuation using the \cite{Inoue2014} model. We model velocity dispersion within our target objects using a Gaussian broadening in velocity space with width $\sigma_\mathrm{disp}$, which is varied with a logarithmic prior.

According to the galaxies' location on the BPT and WHAN diagrams \citep{Baldwin1981,CidFernandes2011}, 6/14 of our sample could have contributions from an AGN in their spectra \citep{Stevenson2026}. Therefore, following \cite{Carnall2023c}, we also test the inclusion of an AGN component, consisting of AGN continuum emission, as well as broad H$\alpha$ and H$\beta$ emission lines. The continuum model follows the broken power law from \cite{VandenBerk2001}, which is described by a break at $\lambda_\mathrm{rest}=5000${\AA} and two power-law indices ($\alpha_{\lambda,<5000}$ and $\alpha_{\lambda,>5000}$). The normalisation of the continuum is parametrized via its flux at rest-frame $5100${\AA} ($f_{5100}$). The broad H$\alpha$ component is modelled with a Gaussian profile, where we fit its normalisation ($f_{\mathrm{H}\alpha, \mathrm{broad}}$) and velocity dispersion ($\sigma_\mathrm{AGN}$). We use the same parameters to model the broad H$\beta$ emission line, but divide its normalisation by $2.86$ assuming case B recombination. 

Although we fit all 14 galaxies using the model including the AGN component, when compared to results from fitting without the AGN component, we see minimal changes in estimated properties (e.g., $\Delta\log_{10}(M_*/\mathrm{M_\odot})<0.02\,$; $\Delta \mathrm{Age}<0.03\,$Gyr). Thus, for the 6 galaxies that lie outside any AGN regions in the WHAN diagram at $>1\sigma$ confidence (45981, 50789, 55410, 55742, 65915 and 117560; see fig. 13 in \citealt{Stevenson2026}), we report results from their fits without the AGN model. For the other 8 objects we report results for the run including the AGN model, however we only observe a significant AGN contribution in 34495, which exhibits a clearly visible broad H$\alpha$ component. Spectrum decomposition following \cite{Krishna2025} also showed a significant AGN contribution in 34495 alone. We therefore conclude that contamination of the continuum emission from our other galaxies by AGN is very weak, if any. Therefore, we consider our results robust against AGN contamination.

Since spectrophotometric calibration has already been performed on the input spectra in Section \ref{sec:excels_reduction}, we do not use the multiplicative Chebyshev calibration polynomial that was employed in \cite{Carnall2024}. Instead, we include an additive Gaussian Process (GP) correlated noise model. This allows for the correction of any remaining minor calibration imperfections, as well as model-data mismatch, and also properly accounts for correlated noise across wavelength bins in the observed spectra \citep{Carnall2019b}. The GP model uses a stochastically driven damped simple harmonic oscillator kernel, as introduced in \cite{Leung2024} and \cite{Leung2025}, implemented through the \texttt{celerite2} python package \citep{Foreman-Mickey2017,Foreman-Mickey2018}. To model potentially underestimated observational uncertainties in our input spectra (e.g., \citealt{Maseda2023}), we also include a multiplicative factor ($s$) on the spectroscopic uncertainties, which is varied with a logarithmic prior. Sampling of the model posterior within \textsc{Bagpipes} is performed using the \textsc{Nautilus} nested sampling algorithm \citep{Lange2023}.

The posterior median models fitted to our spectroscopic data by this process are shown with black lines in Fig. \ref{fig:all_spec}. We also show one fit in more detail for object PRIMER-EXCELS-34495 in Fig. \ref{fig:spec_fit_example}. The top left panel shows the position of the open NIRSpec MSA shutters of the EXCELS G235M observations overlaid on a PRIMER F277W cutout image of the galaxy. The top right panel shows the 2D spectra for the 3 individual gratings observed as part of EXCELS. The central panel shows our 1D spectroscopic extraction (blue), with observed photometry (red points). Over-plotted are the fitted posterior median model spectrum (black), the physical model spectrum (fitted spectrum $-$ GP component, magenta) and the best fit AGN component (red). In the lower panels we also show the residual spectrum and the GP noise model. The bottom left panel shows the full \textit{JWST}+\textit{HST} multi-band photometric data for this galaxy, and the full best fit model spectrum.

\subsubsection{Consistency checks on our full-spectral-fitting methodology}
To test the dependence of our results on our assumed SFH parameterisation, we also repeated our fits using the non-parametric ``continuity'' SFH model \citep{Leja2019a}, following the binning implementation of \cite{Park2024}. We find no significant change in the measured SFHs and subsequent results when switching our fiducial double-power-law model for the continuity non-parametric model (estimates of $t_{50}$, stellar mass and metallicity are within $1\sigma$ of the fiducial estimates, as was also found in \citealt{Carnall2023c}). Thus, we conclude that our findings are not strongly dependent on the SFH model assumed.

\begin{table*}
    \centering
    \caption{Galaxy physical properties derived for our EXCELS sample from the full spectral fitting method described in Section \ref{sec:fitting}. The quantities reported in columns (2), (3), (5), and (9) are as defined in Table \ref{tab:priors}. For the rest, from left to right, they are (1) the PRIMER-EXCELS ID, (4) the star-formation rate averaged over the previous 10 Myr (for measurements with median values $<0.01\,\mathrm{M_\odot\, yr^{-1}}$, we report the $3\sigma$ upper limit; if the upper limit $<0.01\,\mathrm{M_\odot\, yr^{-1}}$, we report this threshold as the upper limit), (6) cosmic time when half of the total stellar mass had been formed, (7) the redshift at which half of the total stellar mass had been formed, (8) the quenching redshift, at which $\mathrm{sSFR}<0.2/t_\mathrm{H}$ is satisfied for the first time, and (10) the time span between when the galaxy had formed 10 per cent and 90 per cent of its total stellar mass. The galaxies' celestial coordinates can be found in Table 1 of \protect\cite{Carnall2024}.}
\renewcommand{\arraystretch}{1.25}
    \setlength{\tabcolsep}{6pt}
    \begin{threeparttable}
    \begin{tabular}{p{0.7cm}p{1.8cm}p{1.6cm}p{1.7cm}p{1.5cm}p{1.3cm}p{1.0cm}p{0.9cm}p{1.2cm}p{1.5cm}}
        \hline
        ID\newline(1) & Redshift\newline(2) & $\log_{10}(M_*/\mathrm{M_\odot})$ (3) & SFR / $\mathrm{M_\odot\, yr^{-1}}$ (4) & $\log_{10}(Z_*/\mathrm{Z_\odot})$ (5) & $t_\mathrm{form}$ / Gyr (6) & $z_\mathrm{form}$\newline(7) & $z_\mathrm{quench}$ (8) & $A_V$ / mag (9) & $\tau_{10-90}$ / Myr (10) \\
        \hline
        34495       & $3.7986\pm0.0002$                & $10.79\pm0.02$                   & $20.77\pm3.68$                   & $\phantom{-}0.45^{+0.06}_{-0.10}$ & $1.43^{+0.02}_{-0.02}$           & $4.2^{+0.1}_{-0.0}$              & --\tnote{*}                               & $1.11^{+0.06}_{-0.06}$           & $112^{+19}_{-11}$                 \\
        39063       & $3.7021\pm0.0002$                & $10.49\pm0.02$                   & $<0.01$                          & $\phantom{-}0.03^{+0.08}_{-0.10}$ & $1.26^{+0.02}_{-0.02}$           & $4.7^{+0.1}_{-0.1}$              & $4.52^{+0.10}_{-0.11}$           & $0.23^{+0.03}_{-0.03}$           & $25^{+30}_{-14}$                  \\
        45981       & $3.0795\pm0.0003$                & $10.59\pm0.02$                   & $<0.01$                          & $\phantom{-}0.16^{+0.10}_{-0.09}$ & $1.02^{+0.08}_{-0.13}$           & $5.5^{+0.6}_{-0.3}$              & $4.78^{+0.49}_{-0.55}$           & $0.67^{+0.09}_{-0.07}$           & $150^{+722}_{-123}$               \\
        50789       & $3.9889\pm0.0002$                & $11.00\pm0.02$                   & $<0.01$                          & $-0.13^{+0.09}_{-0.09}$          & $1.01^{+0.03}_{-0.03}$           & $5.6^{+0.1}_{-0.1}$              & $5.33^{+0.19}_{-0.17}$           & $0.55^{+0.05}_{-0.05}$           & $34^{+59}_{-23}$                  \\
        52467       & $3.2395\pm0.0003$                & $10.50\pm0.02$                   & $2.65\pm0.56$                    & $-0.14^{+0.12}_{-0.10}$          & $1.50^{+0.05}_{-0.04}$           & $4.0^{+0.1}_{-0.1}$              & $3.36^{+0.04}_{-0.04}$           & $0.29^{+0.07}_{-0.06}$           & $211^{+24}_{-21}$                 \\
        55410       & $3.1954\pm0.0002$                & $11.19\pm0.02$                   & $<1.13$                          & $\phantom{-}0.07^{+0.14}_{-0.09}$ & $0.28^{+0.17}_{-0.14}$           & $14.5^{+8.9}_{-4.2}$             & $9.90^{+7.79}_{-3.29}$           & $0.26^{+0.06}_{-0.06}$           & $262^{+287}_{-200}$               \\
        55742\tnote{\textdagger}       & $3.1971\pm0.0005$                & $10.04\pm0.05$                   & $<0.01$                          & $-0.10^{+0.23}_{-0.17}$          & $0.69^{+0.23}_{-0.22}$           & $7.5^{+2.5}_{-1.5}$              & $5.81^{+1.94}_{-0.87}$           & $0.18^{+0.14}_{-0.10}$           & $259^{+477}_{-233}$               \\
        57000       & $3.1946\pm0.0004$                & $10.78\pm0.03$                   & $6.36\pm5.32$                    & $\phantom{-}0.28^{+0.16}_{-0.12}$ & $1.67^{+0.03}_{-0.04}$           & $3.7^{+0.1}_{-0.1}$              & --\tnote{*}                               & $1.26^{+0.07}_{-0.06}$           & $143^{+34}_{-74}$                 \\
        65915       & $4.3599\pm0.0003$                & $10.82\pm0.02$                   & $<0.01$                          & $\phantom{-}0.35^{+0.07}_{-0.25}$ & $0.71^{+0.06}_{-0.10}$           & $7.3^{+0.9}_{-0.4}$              & $6.38^{+0.63}_{-0.61}$           & $0.24^{+0.07}_{-0.06}$           & $125^{+446}_{-104}$               \\
        106260      & $3.9788\pm0.0005$                & $10.80\pm0.04$                   & $<0.01$                          & $\phantom{-}0.44^{+0.08}_{-0.20}$ & $1.19^{+0.03}_{-0.07}$           & $4.9^{+0.2}_{-0.1}$              & $4.69^{+0.18}_{-0.19}$           & $1.66^{+0.09}_{-0.10}$           & $45^{+77}_{-30}$                  \\
        109760      & $4.6220\pm0.0003$                & $11.06\pm0.02$                   & $<0.01$                          & $-0.71^{+0.08}_{-0.08}$          & $0.48^{+0.05}_{-0.04}$           & $9.7^{+0.7}_{-0.7}$              & $6.80^{+0.74}_{-0.55}$           & $0.95^{+0.07}_{-0.07}$           & $522^{+121}_{-253}$               \\
        112990\tnote{\textdagger}      & $3.8924\pm0.0012$                & $10.82\pm0.05$                   & $70.93\pm19.80$                  & $-0.33^{+0.21}_{-0.28}$          & $0.84^{+0.16}_{-0.08}$           & $6.4^{+0.5}_{-0.8}$              & --\tnote{*}                               & $1.97^{+0.15}_{-0.14}$           & $1151^{+78}_{-243}$               \\
        113667      & $3.9725\pm0.0002$                & $10.58\pm0.02$                   & $<0.01$                          & $-0.87^{+0.10}_{-0.14}$          & $1.02^{+0.03}_{-0.03}$           & $5.5^{+0.1}_{-0.1}$              & $5.32^{+0.17}_{-0.16}$           & $0.10^{+0.07}_{-0.05}$           & $31^{+43}_{-20}$                  \\
        117560      & $4.6200\pm0.0003$                & $11.05\pm0.02$                   & $<0.13$                          & $\phantom{-}0.21^{+0.09}_{-0.13}$ & $0.62^{+0.05}_{-0.05}$           & $8.1^{+0.5}_{-0.4}$              & $7.15^{+0.66}_{-0.81}$           & $0.42^{+0.06}_{-0.06}$           & $113^{+236}_{-95}$                \\
        \hline
    \end{tabular}
    \begin{tablenotes}
        \item [\textdagger] We obtain very large posterior uncertainties in most parameters, due to the low SNRs and poor Balmer line coverage of these EXCELS spectra. These two objects are therefore removed from all further analysis.
        \item [*] No measured $z_\mathrm{quench}$ value because the fitted star-formation history does not satisfy the quenched criterion of $\mathrm{sSFR}<0.2/t_\mathrm{H}$ at any point.
    \end{tablenotes}
    \end{threeparttable}
    \label{tab:main_results}
\end{table*}

The GP model corrections on our best-fit spectral models typically have a magnitude $<4$ per cent of the input spectrum, and are hence typically smaller than the observational uncertainties (see the bottom right panel in Fig. \ref{fig:spec_fit_example}). To investigate the impact of our GP model, we repeated our fits, firstly with a 38th order multiplicative Chebyshev polynomial instead of the GP noise model, then again with both the polynomial and GP models. These tests produce estimated SFHs and galaxy bulk properties (e.g., stellar mass, mass-weighted age) within the $1\sigma$ uncertainty region of our fiducial model. Therefore, we conclude that our choice of the GP model over the polynomial method does not significantly impact our results.

\section{Results} \label{sec:results}

\begin{figure}
    \centering
    \includegraphics[width=\columnwidth]{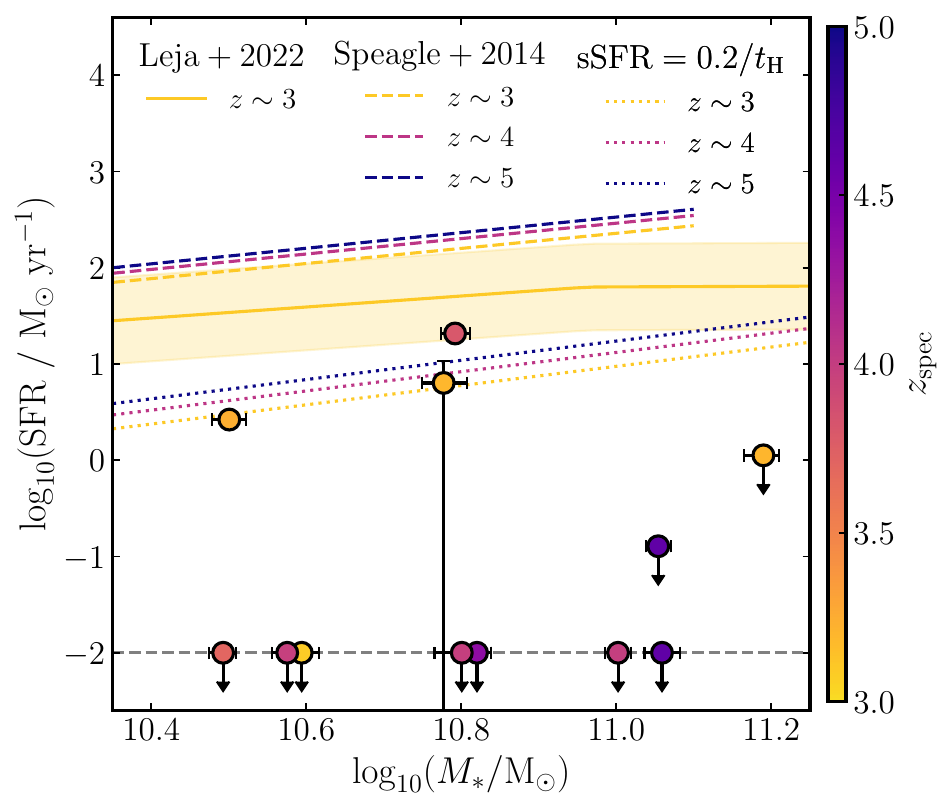}
    \caption{The stellar masses of our sample of 12 massive quiescent galaxies plotted against their SFRs (averaged over 10 Myr), with both quantities measured from full spectral fitting. For data points with posterior median $\log_{10}(\mathrm{SFR/M_\odot\,yr^{-1}})<-2$, we show $3\sigma$ upper limits. We also set a minimum SFR upper limit at $\log_{10}(\mathrm{SFR/M_\odot\,yr^{-1}})<-2$ (grey dashed lines) to limit the vertical dynamic range of the figure. The symbols are coloured according to their observed redshifts. We show the SFMS from \protect\cite{Leja2022} at $z\sim3$ with a solid line and shaded region. We show the SFMS from \protect\cite{Speagle2014} at $z\sim3$, $z\sim4$ and $z\sim5$ with dashed lines. We also mark the threshold $\mathrm{sSFR}=0.2/t_\mathrm{H}$ at $z\sim3$, $z\sim4$ and $z\sim5$ with dotted lines.}
    \label{fig:SFMS}
\end{figure}

The properties of the 14 EXCELS $z>3$ massive quiescent galaxies measured via the full-spectral-fitting approach laid out in Section \ref{sec:fitting} are summarised in Table \ref{tab:main_results}. Their celestial coordinates can be found in Table 1 of \cite{Carnall2024}. Two galaxies with the noisiest EXCELS spectra and poor coverage of the Balmer lines (55742 and 112990; SNR per {\AA} $<3$ averaged over $5100<\lambda_\mathrm{rest}<6400\,${\AA}) returned fitted parameters and SFHs with very large uncertainties, and we therefore exclude these galaxies from all further analysis. Our results are summarised in Figs \ref{fig:SFMS} and \ref{fig:simple_results}, and compared with results from the literature in Fig. \ref{fig:results}. We show a gallery of our fitted SFH posteriors in Fig. \ref{fig:all_sfh}.

\subsection{Current star-formation rates}\label{sec:sfr_results}

For 9/12 of the galaxies with spectroscopic data of a high enough quality to obtain robust results we recover $\mathrm{SFR}<1\ \mathrm{M_\odot\, yr^{-1}}$ at the time of observation. Although the other three (34495, 52467 and 57000) appear not to have fully shut down star formation when observed, they all display sharp recent declines in SFR (see Fig. \ref{fig:all_sfh}), indicating that they are all rapidly quenching and will soon reach quiescence. Whilst two of our objects (34495 and 57000) fall above the $\mathrm{sSFR}<0.2/t_\mathrm{H}$ threshold used by \cite{Carnall2024} to select the EXCELS sample, 57000 is within $1\sigma$ of this threshold, and 34495 $(z=3.78)$ is just 0.4 dex higher. The latter object also has a measured SFR significantly lower than the $z\sim4$ SFMS from \cite{Speagle2014}, and is included as a quiescent galaxy in the DeepDive sample presented by \cite{Hamadouche2026}. Therefore, even though 34495 and 57000 do not strictly meet the quiescence threshold used by \cite{Carnall2024}, we retain all 12 galaxies in our sample.

From the final 2 objects for which we do not obtain robust results via our full-spectral-fitting methodology, 112990 shows some signs of a higher star-formation rate, consistent with the spectrum showing detectable [O\,\textsc{iii}] line emission in Fig. \ref{fig:all_spec}. However, the poor quality of the spectrum makes it challenging to constrain the SFR precisely (SNR per {\AA} $=2.74$ and missing the Balmer break region). A contamination rate of $1-2$ objects from our sample of 14 is consistent with the $8\pm3$ per cent spectroscopic contamination rate derived by \cite{Stevenson2026} for a larger photometric sample selected via the same process.

\subsection{Star-formation histories}

\subsubsection{The stellar mass vs stellar age relationship}

\begin{figure*}
    \centering
    \includegraphics[width=0.9\textwidth]{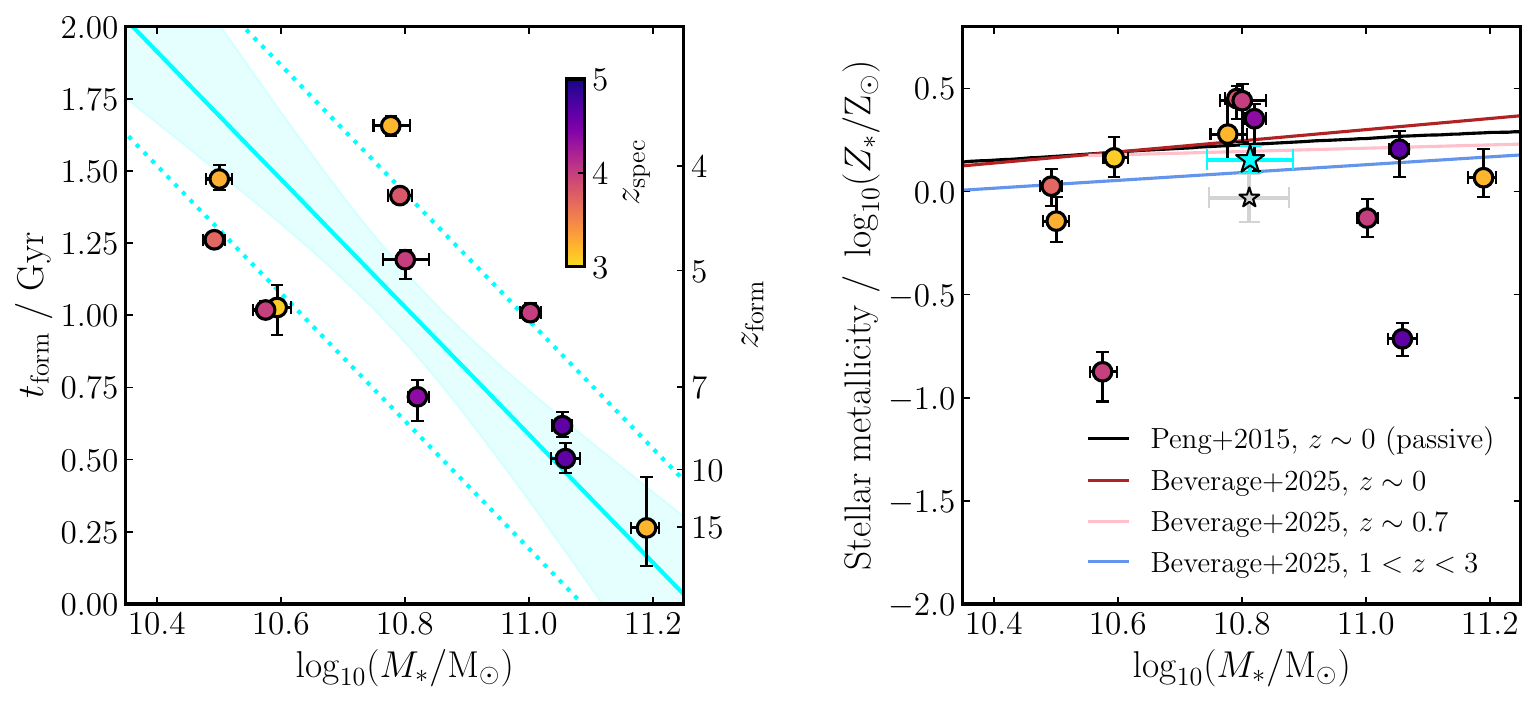}
    \caption{Relationships between stellar population properties for the 12 EXCELS $z>3$ massive quiescent galaxies for which we obtain robust results. \textbf{Left}: Age of the Universe at which 50 per cent of the stellar mass in each galaxy had formed ($t_\mathrm{form}$) as a function of stellar mass. Each galaxy is coloured according to its observed redshift ($z_\mathrm{spec}$). We fit a straight line with intrinsic scatter to the tight correlation visible, which is represented by the light blue line and shaded region (Equation \ref{eq:age_vs_mass}). The dotted light blue lines correspond to the $\pm1\sigma$ intrinsic scatter we measure for the best-fit relation. \textbf{Right}: Stellar metallicity as a function of stellar mass, coloured in the same way as the left panel. The grey star shows the inverse-variance-weighted mean of our sample, with error bars showing the standard error on the mean. The cyan star shows the inverse-variance-weighted mean excluding the two very low metallicity objects (see Section \ref{sec:metallicity}). For comparison, we also show the stellar mass-metallicity relations of passive/quiescent galaxies from \protect\cite{Peng2015} at $z\sim0$ (light-weighted, black line) and \protect\cite{Beverage2025} at $z\sim0$ (red line), $z\sim0.7$ (pink line) and $1<z<3$ (blue line).}
    \label{fig:simple_results}
\end{figure*}

In the left panel of Fig. \ref{fig:simple_results}, we plot the estimated stellar masses of the 12 massive quiescent galaxies against the cosmic times at which we estimate that half their stellar mass had formed ($t_\mathrm{form}$, measured forwards from the Big Bang). Objects are coloured according to their observed redshifts. A tight negative relation is observed, where more massive galaxies formed the bulk of their stellar masses earlier than less massive galaxies. As discussed in Section \ref{sec:intro}, this is widely observed at lower redshift, and is known as ``downsizing'' or sometimes ``archaeological downsizing'' \citep[e.g.,][]{Cowie1996,Perez-Gonzalez2008,Thomas2010}.

Similar to earlier works at lower redshift \citep[e.g.,][]{Gallazzi2014,Carnall2019b,Hamadouche2023}, we follow the methods detailed in \cite{Hogg2010} to fit a linear relationship with intrinsic scatter in the vertical direction. For the mean relationship, we find
\begin{equation}\label{eq:age_vs_mass}
    \bigg(\frac{t_\mathrm{form}}{\mathrm{Gyr}}\bigg) = 0.59^{+0.16}_{-0.23} -2.21^{+0.56}_{-0.93}\,\log_{10}\bigg(\frac{M_*}{10^{11}\mathrm{M_\odot}}\bigg) \, ,
\end{equation}
with an intrinsic scatter of $0.40^{+0.21}_{-0.11}$ Gyr. The mean relationship is shown in Fig. \ref{fig:simple_results} as a cyan line, while its $1\sigma$ confidence region is marked with the shaded region. The pair of dotted cyan lines indicate the mean relationship $\pm$ intrinsic scatter. Both jackknife and bootstrap resampling tests yield consistent results to that reported above, confirming that despite the small sample size, the observed relation is not driven by one or two outlying galaxies.

Notably, the three most massive galaxies in our sample formed extremely early. Our measured SFH suggests that the most massive galaxy in our sample, PRIMER-EXCELS-55410 (otherwise known as ZF-UDS-7329), had formed half of its total stellar mass by $t_\mathrm{form}=0.28^{+0.17}_{-0.14}\,$Gyr ($z_\mathrm{form}=14.5^{+9.0}_{-4.2}$). This is consistent with the red continuum shape, strong 4000{\AA} break and weak Balmer absorption lines that can be seen for this object in Fig. \ref{fig:all_spec}. This result is in good agreement with the results of previous studies that have focused on this object \citep{Glazebrook2023,Carnall2024,Turner2025,Nanayakkara2025}. It is also worth noting that the formation redshifts we infer for PRIMER-EXCELS-109760 and 117560 at $z=4.62$ are consistent with the results of \cite{Carnall2024}.

\begin{figure*}
    \centering
    \includegraphics[width=\textwidth]{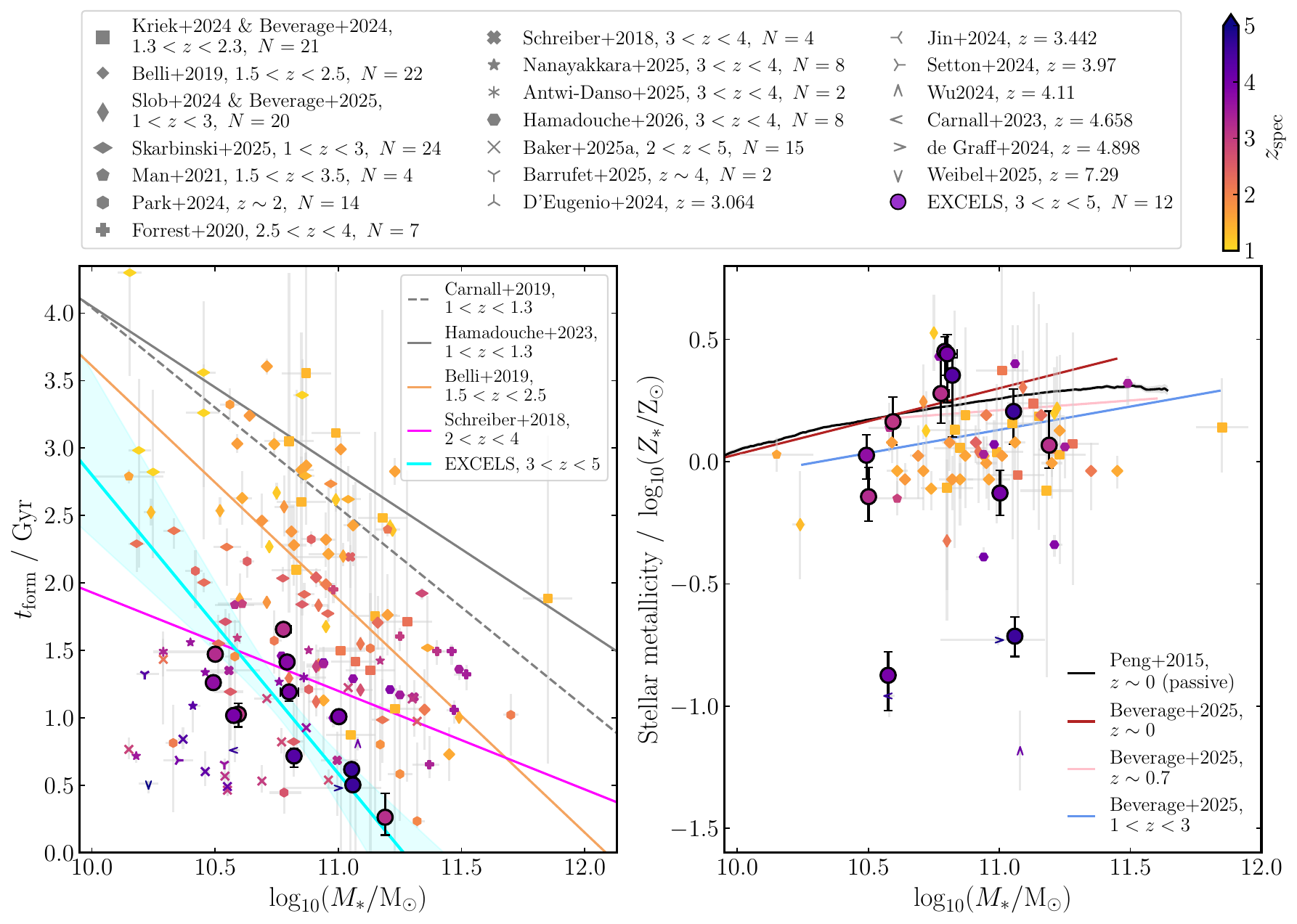}
    \caption{The stellar population properties for our sample of massive quiescent galaxies, as shown in Fig. \ref{fig:simple_results}, placed in literature context. In both panels, we have included individual massive quiescent galaxies reported in the literature at $z>1$ (see Sections \ref{sec:downsizing} and \ref{sec:metallicity} for full lists of references). \textbf{Left}: Age of the Universe at which 50 per cent of the stellar mass in each galaxy had formed ($t_\mathrm{form}$) as a function of stellar mass. The cyan line and shaded region is our best-fit relation to the EXCELS sample, as shown in Fig. \ref{fig:simple_results}. We also plot the mean relations from \protect\cite{Hamadouche2023} (grey solid line) and \protect\cite{Carnall2019b} (grey dashed line) at $1<z<1.3$. In addition, we plot the mean relations measured by \protect\cite{Hamadouche2023} using samples from \protect\cite{Belli2019} at $1.5<z<2.5$ (orange line) and \protect\cite{Schreiber2018} at $3<z<4$ (magenta line). Our galaxies follow a mass-age sequence, with a slope consistent with those reported for spectroscopic samples at $1<z<2.5$. The \protect\cite{Schreiber2018} relationship is shallower, probably due to their SFHs being derived purely from photometric data. \textbf{Right}: Stellar metallicity as a function of stellar mass. We again show the mean relationships plotted in Fig. \ref{fig:simple_results}, as well as individual-object results from a range of studies. We see no clear mass-metallicity relation in our relatively small sample. However, excluding our 3 lowest-metallicity objects, our results are consistent with the literature at $1 < z < 3$. Note that compared to the previous figures, the colour map has been expanded to cover the redshift range $1 < z < 5$.}
    \label{fig:results}
\end{figure*}

\subsubsection{Archaeological downsizing and its evolution since $z\sim5$} \label{sec:downsizing}

Our results in Fig. \ref{fig:simple_results} provide a clear spectroscopic confirmation that the archaeological downsizing trend is already in place at $3 < z < 5$. We next compare the fitted $t_\mathrm{form}$-mass relation from our sample with other results from the literature at $z>1$ and $\log_{10}(M_*/\mathrm{M_\odot})>10$ in the left panel of Fig. \ref{fig:results}. The literature galaxies are obtained from \cite{Schreiber2018}\footnote{Originally 12 galaxies. Three removed due to emission lines contaminating fits to observed photometry following \cite{Schreiber2018}; five removed due to overlapping with the \cite{Nanayakkara2025} and \cite{Hamadouche2026} samples.}, \cite{Belli2019}\footnote{Originally 23 galaxies. One removed due to overlapping with the \cite{Park2024} sample.}, \cite{Forrest2020a}\footnote{Originally 8 galaxies. One removed due to overlapping with the \cite{Hamadouche2026} sample.}, \cite{Man2021}, \cite{Carnall2023c}, \cite{Kriek2024}, \cite{Slob2024}, \cite{Park2024}, \cite{Jin2024}, \cite{Setton2024}, \cite{Nanayakkara2025}\footnote{Originally 17 galaxies. Two removed due to missing Balmer region in observed spectra; seven removed due to overlapping with the EXELS and \cite{Hamadouche2026} samples.}, \cite{DEugenio2024}, \cite{Antwi-Danso2025}, \cite{Baker2025b}\footnote{Originally 18 galaxies. Three removed due to overlapping with the \cite{Barrufet2025} sample and GS-9029 in \cite{Carnall2023c}.}, \cite{Barrufet2025}, \cite{Wu2025}, \cite{deGraaff2025}, \cite{Weibel2025}, \cite{Skarbinski2026}, and \cite{Hamadouche2026}\footnote{Originally 10 galaxies. Two removed due to overlapping with the EXCELS sample}.

We begin by considering the work of \cite{Schreiber2018}, who analysed ground-based spectroscopy for a sample of massive quiescent galaxies at $3 < z < 4$. Although probing massive quiescent galaxies at similar redshifts, our $t_\mathrm{form}$-mass relation appears to be in tension with that obtained by \cite{Schreiber2018} (we plot the relationship derived by \citealt{Hamadouche2023} from the \citealt{Schreiber2018} results), who measured a flatter slope of $-0.73^{+0.64}_{-0.69}\,$Gyr per dex in stellar mass, compared with our $-2.21^{+0.56}_{-0.93}$ Gyr per dex value. However, the stellar masses and ages in the \cite{Schreiber2018} sample were measured from fitting photometry alone given spectroscopic redshifts. This approach suffers from increased uncertainties and can lead to a flattening of the $t_\mathrm{form}$-mass relation (see section 6.1 of \citealt{Carnall2019b}). Additionally, due to the small sample sizes in both our study ($N=12$) and \cite{Schreiber2018} ($N=12$), the tension is not highly significant, at only $1.66\sigma$. 

Many of the galaxies observed by \cite{Schreiber2018} received significantly deeper JWST NIRSpec $R\sim1000$ spectra in the G235M grating as part of the DeepDive survey \citep{Ito2025b}. \cite{Hamadouche2026} (flat-topped hexagons in Fig. \ref{fig:results}) recently reported the SFHs of 10 massive quiescent galaxies at $3<z<4$ from this sample. Excluding two galaxies from their sample that overlap with the EXCELS sample (DD111 is PRIMER-EXCELS-34495; DD80 is PRIMER-EXCELS-50789), their galaxies are generally more massive and exhibit younger stellar ages (higher $t_\mathrm{form}$) compared to our results. This difference is likely to be due to selection effects, since the DeepDive galaxies were selected from ground-based photometry, which recent \textit{JWST} studies have found to miss a significant portion of fainter and redder quiescent objects at these redshifts (e.g., \citealt{Carnall2023b}). Therefore, DeepDive would be expected to be biased towards brighter and bluer galaxies, and thus younger stellar ages, as is observed. Since the EXCELS sample was selected from \textit{JWST} PRIMER photometry, our sample of massive quiescent galaxies is more representative of the true underlying population at $3<z<5$.

Comparing to lower-redshift results, the slope of our $t_\mathrm{form}$-mass relation shows reasonable agreement with those measured at the cosmic noon epoch $(1 \lesssim z \lesssim 3)$ from spectroscopic data. Using ultra-deep spectroscopy for 114 massive quiescent galaxies from the VANDELS survey, \cite{Hamadouche2023} found quiescent galaxies at $1<z<1.3$ exhibit an age-mass slope of $-1.20^{+0.28}_{-0.27}\,$Gyr per dex in stellar mass. \cite{Hamadouche2023} also measured an age-mass slope for the 23 quiescent galaxies reported by \cite{Belli2019} at $1.5 < z < 2.5$, derived from Keck-MOSFIRE spectroscopy. Their result is $-1.73\pm0.40\,$Gyr per dex in stellar mass. The slopes from both samples are within $1\sigma$ of our measured slope in Equation \ref{eq:age_vs_mass}. It is possible that the slope steepens towards higher redshift, but the large uncertainty on this work's measured slope prevents a clear conclusion from being drawn.

The analysis of \cite{Carnall2019b} (an earlier version of that presented by \citealt{Hamadouche2023}) also concluded that a $\simeq1.5$ Gyr per dex slope, within $1\sigma$ uncertainty of our measured slope, appears to be consistent with the results of \cite{Gallazzi2005} at $z\simeq0.1$ and \cite{Gallazzi2014} at $z\simeq0.7$, however substantial methodological differences (e.g., the use of light-weighted, rather than mass-weighted ages) makes this comparison more challenging.

The broadly parallel nature of these age-mass relationships across $0 < z < 5$ indicates that, with decreasing redshift, the increase in mean quiescent galaxy formation time is largely independent of stellar mass. Given that our results show clear evidence for the archaeological downsizing trend already being in place by $z\sim4$, only $\sim1.5\,$Gyr after the Big Bang, some of the proposed drivers of downsizing discussed in Section \ref{sec:intro}, such as environmental effects and dry mergers, are unlikely to have had enough time to make a significant contribution (but see \citealt{Ito2025a}). 

\begin{figure}
    \centering
    \includegraphics[width=\columnwidth]{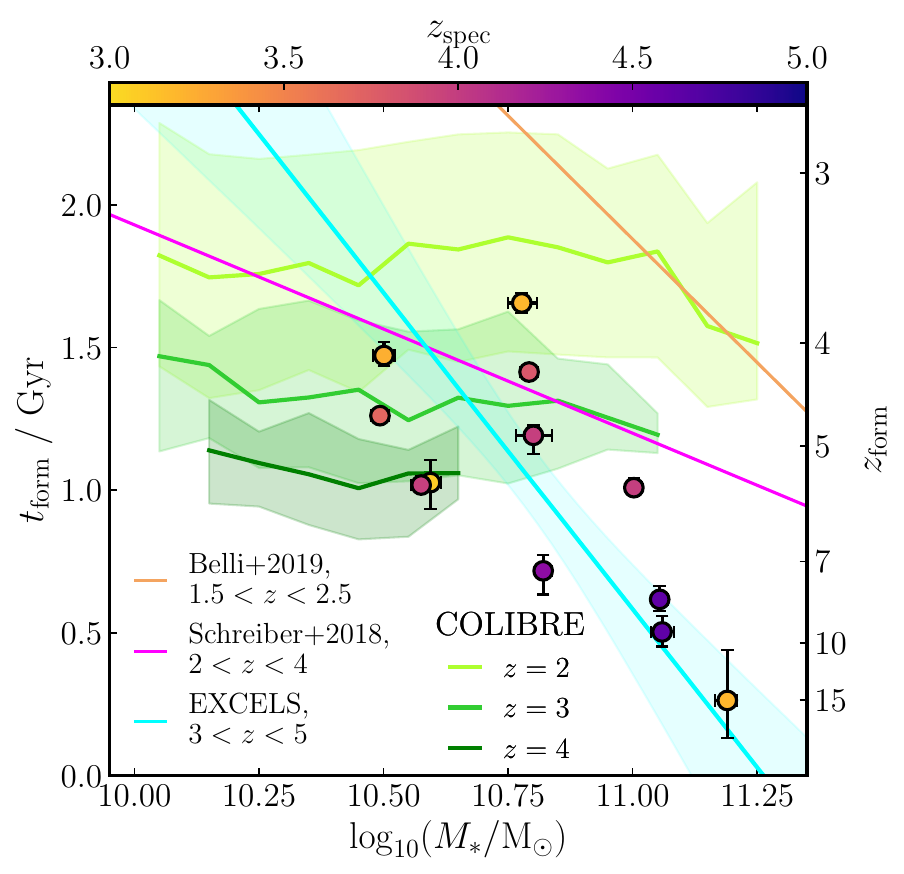}
    \caption{A comparison of observational age-mass relationships for massive quiescent galaxies to predictions from simulations. The vertical axis shows the age of the Universe at which galaxies formed the 50th percentile of their stellar mass. We plot the individual-galaxy EXCELS results (circles), our best fit age-mass relationship (cyan) and the relationships measured by \protect\cite{Hamadouche2023} from the lower redshift samples of \protect\cite{Belli2019} (orange) and \protect\cite{Schreiber2018} (magenta) in similar styles to Fig. \ref{fig:results}. The \protect\cite{Chandro-Gomez2025} relationships predicted by the COLIBRE simulation at $z=2$, $z=3$ and $z=4$ are shown in shades of green, with shaded regions indicating the 16th to 84th percentile ranges.}
    \label{fig:age_sim}
\end{figure}

Finally, it is worth noting that our sample of massive quiescent galaxies at $3<z<5$ overlaps in the left panel of Fig. \ref{fig:results} with the galaxies at $z\simeq1$ that formed at the earliest times, suggesting that a fraction of massive quiescent galaxies at $3<z<5$ will remain quiescent for $>3\,$Gyr (and perhaps much longer), becoming ancient relics by $z\lesssim1$ (e.g., \citealt{Ferre-Mateu2017, Spiniello2021}).

\subsubsection{Archaeological downsizing in simulations}

The archaeological downsizing trend is also of interest because it can be readily compared with predictions from cosmological simulations (e.g., \citealt{Nelson2018, Carnall2019b, Lovell2023b}). This provides a useful additional constraint to number-density comparisons, being strongly dependent on the physical process that gives rise to quenching. In Fig. \ref{fig:age_sim}, we compare our observed age-mass relationship with those predicted from the COLIBRE cosmological hydrodynamic simulation \citep{Schaye2025,Chaikin2025}, taken from \cite{Chandro-Gomez2025}, for objects selected in the same way as our observational sample. COLIBRE is notable for accurately reproducing the number density of massive quiescent galaxies at $3 < z < 5$ \citep{Chandro-Gomez2025}, whereas many other simulations under-predict the abundance of such objects (e.g., \citealt{Lagos2025,Stevenson2026}).

It can be seen from Fig. \ref{fig:age_sim} that COLIBRE broadly predicts flat age-mass relationships at all redshift bins between $z=2$ and $z=4$, but the simulation suffers from a lack of statistics at the high-mass end at the highest redshifts. In contrast, our observed relation and those from \cite{Schreiber2018} and \cite{Belli2019} all exhibit steeper slopes. We therefore report tentative evidence that, although COLIBRE is broadly able to reproduced the observed number densities of massive quiescent galaxies at $z>2$, there is some evidence that the observed stellar ages are not yet well reproduced. Given that COLIBRE reproduces the observed number densities of massive quiescent galaxies at $3<z<5$, but the galaxies predicted might be too young compared to observations, the rates of both quenching and rejuvenation in the simulations might be too high. It has been previously noted that other simulations also struggle to reproduce the oldest $z>3$ quiescent galaxies, instead producing exclusively young objects (e.g., \citealt{Hartley2023,Weller2025}). 

\subsubsection{Star-formation-history shapes}

We show the fitted SFHs of all 12 EXCELS galaxies for which we obtain a good fit in Fig \ref{fig:all_sfh}. It can be seen that these objects generally experienced extremely rapid assembly of their stellar mass. The assembly duration, which we measure via the time span between when the galaxy had formed 10 per cent and 90 per cent of its total stellar mass ($\tau_{10-90}$), has a mean value of $99\pm10\,$Myr. Only 1/12 of our  galaxies, 109760, has a posterior median assembly duration, $\tau_{10-90}>300\,$Myr. Such rapid assembly requires peak SFRs of several hundred solar masses per year, a level that is comparable to the SFRs of the most extreme submillimetre galaxies at the redshifts our objects formed \citep[e.g.,][]{Michalowski2017, Liu2025, Bing2025}.

\subsubsection{Implications for quiescent galaxies at higher redshifts}
The age-mass relationship we report in Equation \ref{eq:age_vs_mass} has interesting implications for extremely early massive quiescent galaxies, such as the object RUBIES-UDS-QG-z7 reported by \cite{Weibel2025} at $z=7.3$, which appears to have formed its stellar mass of log$_{10}(M_*/\mathrm{M_\odot})\simeq10.2$ at $z\simeq8-9$. Whilst our sample is only mass complete at $\log_{10}(M_*/\mathrm{M_\odot})>10.4$ (above the mass of RUBIES-UDS-QG-z7), and probes a smaller area compared to the area from which RUBIES-UDS-QG-z7 was selected, the clear implication of the tight age-mass correlation we observe is that massive quiescent galaxies at $3 < z < 5$ with $\log_{10}(M_*/\mathrm{M_\odot})\gtrsim10$ have formed and quenched very recently, consistent with recent results from photometric data \citep[e.g.,][]{Merlin2025}. The only objects we find in our sample with similar $t_\mathrm{form}$ to RUBIES-UDS-QG-z7 all have $\log_{10}(M_*/\mathrm{M_\odot})\gtrsim11$. This implies that RUBIES-UDS-QG-z7 will likely remain quiescent for only a short duration, before rejuvenating.

\subsection{Stellar metallicities}\label{sec:metallicity}

In the right panel of Fig. \ref{fig:simple_results} we plot our measured stellar masses against our measured stellar metallicities. We also show several median relationships from the literature. The black line shows the result of \cite{Peng2015}, measured from local Sloan Digital Sky Survey (SDSS) quiescent galaxies. The other three relationships are those reported by \cite{Beverage2025}. Their $z\sim0$ line, shown in red, is derived using individual galaxy results from \cite{Zhuang2023}, again based on SDSS data. The $z\sim0.7$ line, shown in pink, is derived from the LEGA-C survey results of \cite{Beverage2023}. The blue $1 < z < 3$ line is derived from SUSPENSE survey data \citep{Slob2024}.

We observe larger scatter in our sample compared to galaxies at lower redshifts, in particular towards far lower metallicities. The inverse-variance-weighted mean metallicity we measure for our sample, marked with a grey star in Fig. \ref{fig:simple_results}, is $\log_{10}(Z_{*,\mathrm{mean}}/\mathrm{Z_\odot})=-0.03\pm0.12$. This lies below the $1<z<3$ relation from \cite{Beverage2025}, however this is primarily driven by the two outlying metal-poor galaxies in our sample. If we exclude these two galaxies, the distribution of the remaining objects in our sample appears to agree well with the \cite{Beverage2025} $1 < z < 3$ relation. The inverse-variance-weighted mean metallicity excluding the two low-metallicity objects, which is shown with a cyan star in Fig. \ref{fig:simple_results}, is $\log_{10}(Z_{*,\mathrm{mean}}/\mathrm{Z_\odot})=0.15\pm0.06$.

The two low-metallicity galaxies in our sample have moderately young stellar populations ($\approx500\,$Myr since quenching), which is also true for the literature examples cited above. The spectrum of 113667 in particular has a clear triangular shape and very deep Balmer absorption lines, closely resembling that of an A-type star. In this context, it seems plausible that these anomalously low stellar metallicity results are instead due to an inadequacy of current low metallicity stellar population models in this relatively poorly explored age regime. We have tested re-fitting these three objects whilst requiring higher stellar metallicities, and confirm that, whilst this results in a lower quality of fit, the stellar ages of these objects are not strongly affected, and the effect on the relationship presented in Equation \ref{eq:age_vs_mass} is minimal.

Next, in the right panel of Fig. \ref{fig:results}, we place our results from Fig.~\ref{fig:simple_results} in the context of a large number of individual stellar metallicity measurements for spectroscopically observed $\log_{10}(M_*/\mathrm{M_\odot})>10$ quiescent galaxies at $z>1$ from the literature. The results are obtained from \cite{Belli2019}, \cite{Man2021}, \cite{Carnall2023c}, \cite{Beverage2024,Beverage2025}\footnote{\cite{Beverage2024} presents the same sample as \cite{Kriek2024}, while \cite{Beverage2025} presents the same sample as \cite{Slob2024}.}, \cite{Wu2025}, \cite{deGraaff2025}, and \cite{Hamadouche2026}.
In general, our stellar metallicity results appear to support a picture in which the quiescent stellar mass-metallicity relation shifts only very modestly towards lower metallicities at earlier times.

The two galaxies with significantly lower metallicity estimates, PRIMER-EXCELS-109760 and 113667, are consistent with the strongly sub-solar metallicities that have recently been reported for several massive quiescent galaxies observed at $z>4$ \citep{Carnall2023c,deGraaff2025,Wu2025}. Combined with the examples in the literature, these galaxies might represent a new type of evolutionary path for early massive quiescent galaxies that rapidly formed and quenched while maintaining substantially sub-solar chemical abundances. However, it is challenging to explain why no such objects appear to be found at lower redshift (though it is possible these galaxies will later become more metal-rich by rejuvenating or accreting more metal-rich stars through mergers).

When comparing our results to the literature relationships shown, it is important to note that they have been measured via different methods. The \cite{Peng2015} relation is based on values first measured by \cite{Gallazzi2005}, who measured light-weighted metallicities from a selection of Lick indices, spectral breaks and absorption features. The difference between light-weighted metallicities and mass-weighted metallicities in quiescent galaxies is however estimated to be $<0.1\,$dex \citep{Trussler2020}.

The \cite{Bruzual2003} models used in our primary fitting approach assume scaled-solar elemental abundances, whereas high-redshift galaxies with rapid formation timescales are expected to be significantly $\alpha$-enhanced (e.g., \citealt{Thomas2005, Kriek2016, Kobayashi2020}). In \cite{Beverage2025}, the authors compared stellar metallicities measured assuming scaled-solar chemical abundances against measurements from fitting the abundances of individual elements separately (albeit assuming a simpler treatment of SFH, dust attenuation and nebular emission). The authors found that metallicity values from full spectral fitting could be up to $\simeq0.7$ dex lower than values derived from measuring individual abundances (median offset $\sim0.4$ dex, see the middle panel of their fig. 6). This could also provide an explanation for the three very low stellar metallicity objects in our sample, however unfortunately the \cite{Conroy2018} stellar population models used by \cite{Beverage2025} are only available for ages $>1$ Gyr. 

The general lack of available models for $\alpha$-enhanced stellar populations covering broad age and wavelength ranges has long limited progress in this area, however several groups have recently developed new sets of $\alpha$-enhanced stellar population models (e.g., \citealt{Knowles2021, Knowles2023, Byrne2025, Park2025}). We therefore explore fitting our sample with stellar models that allow flexibility in elemental abundance patterns in Section \ref{sec:alpha}.

\subsection{AGN contributions and black-hole masses}
Only one galaxy in our sample, 34495, is best fitted with a significant AGN contribution. It is also the only galaxy where the [Fe \textsc{vii}] $\lambda4893$ emission line associated with AGN is detected \citep{Nussbaumer1970,Reefe2023}. We estimate a broad H$\alpha$ flux of $f_{\mathrm{H}\alpha, \mathrm{broad}}=1.5\pm0.3 \times 10^{-17}$ $\mathrm{erg\, s^{-1}\, cm^{-2}}$ and full width at half maximum (FWHM) of $1000^{+1900}_{-1800}\,\mathrm{km\,s^{-1}}$. This broad-line width is similar to that measured for the quiescent galaxy GS-9209 at higher redshift \citep{Carnall2023c}, as well as some $z\approx6$ quasars \citep[e.g.,][]{Chehade2018,Onoue2019}. 

From our measured broad H$\alpha$ flux for 34495, we use the relation in equation 6 of \cite{Greene2005} to estimate a black-hole mass of $\log_{10}(M_\mathrm{BH}/\mathrm{M_\odot})=8.5\pm0.2\,$, which corresponds to a black-hole-to-stellar mass ratio of $M_\mathrm{BH}/M_*=0.6^{+0.3}_{-0.2}$ per cent. Interestingly, this mass ratio is consistent with the black-hole-to-bulge-mass ratios measured from local early type galaxies by \cite{Kormendy2013}, and does not follow the trend towards heightened mass ratios recently reported for many $z>4$ galaxies \citep[e.g.,][]{Pacucci2024,Maiolino2024}. This is consistent with 34495 going on to form the bulge component of a local galaxy with little further evolution.

The black-hole-to-stellar mass ratio we measure for 34495 is also lower than the ratio recently derived for the similar object GS-9209 \citep{Carnall2023c} at $z=4.66$, which has a higher black-hole-to-stellar mass ratio of $1.1^{+0.4}_{-0.3}$ per cent. Whilst this relationship is known to have considerable scatter, and estimates at these redshifts are still scarce, our new result for 34495 combined with the previous GS-9209 result suggests that the average black-hole-to-stellar mass ratio for massive quiescent galaxies at $z\gtrsim4$ may be similar to the 0.8 per cent average black-hole-to-bulge mass ratio derived by \cite{McLure2006} for the most massive elliptical galaxies at $z=2$.

\section{Alpha enhancement} \label{sec:alpha}
Alpha ($\alpha$) elements are elements that are mainly produced through the $\alpha$ process in stellar nucleosynthesis, such as C, O, Ne, Mg, Ca and Si. They are released into the interstellar medium (ISM) in greater quantities through Type II supernovae, while Fe-peak elements are released mainly through Type Ia supernovae \citep[e.g.,][]{Maiolino2019}. The different onset timescales of these two processes after a starburst (Type II $\approx10\,$Myr; Type Ia $\approx1\,$Gyr, see \citealt{Maoz2012}) mean that $\alpha$-abundance is a powerful tracer of the duration of past star formation \citep[e.g.,][]{Thomas2010}. 
Our fiducial results presented in Section \ref{sec:results} assume scaled-solar stellar abundances, however early galaxies have been shown to be more $\alpha$-abundant compared to other elements than is true for the Sun, or `$\alpha$-enhanced' \citep{Steidel2016,Topping2020,Cullen2021,Zhuang2023,Stanton2024,Beverage2025,Shapley2025}. Assuming scaled-solar $\alpha$-abundances could potentially bias our stellar age and metallicity results due to degeneracies between these parameters \citep{Vazdekis2015,Choi2019}. Hence, it is important to explore $\alpha$-enhancement in our galaxies and its impact on the quality of our measured galaxy properties.

\subsection{Variable $\alpha$-abundance fitting implementation}\label{sec:alpha_method}

Metal absorption features in the rest-frame optical typically increase in strength with increasing stellar population age (as well as metallicity), therefore these features are not typically visible in the atmospheres of hot O, B and A-type stars. Thus, in this section we use only galaxies for which the posterior median time since half of their stellar mass formed exceeds 500 Myr (i.e., $t_\mathrm{H}(z)-t_\mathrm{form} > 500\,$Myr) in our fiducial \cite{Bruzual2003} fitting run (see Table \ref{tab:main_results}).

To ensure faint metal absorption features in the spectra are detectable so that [$\alpha$/Fe] can be constrained, we also impose a threshold on SNR, requiring SNR per {\AA} $>6$ averaged over $5100 < \lambda_\mathrm{rest}<6400\,${\AA}. 
For 65915, the Mg absorption complex at $\lambda_\mathrm{rest}\approx5170$\,{\AA} was not observed due to a detector gap. As this feature is a key indicator of $\alpha$-abundance \citep{Thomas2003,Byrne2022,Knowles2023,Park2025}, we also exclude this galaxy. Therefore, in this section we repeat our full spectral fitting with variable [$\alpha$/Fe] abundance ($-0.2$ to $0.6$ dex) for 6 galaxies: 45981, 50789, 55410, 109760, 113667 and 117560.

We conduct 5 separate fitting runs allowing $\alpha$-abundance to vary, described in the following sub-sections, using both the BPASS and sMILES models, which we implement within the \textsc{Bagpipes} code, and the sMILES and \cite{Conroy2018} models using the \texttt{alf}-$\alpha$ code. A summary of these 5 fitting approaches is provided in Table \ref{tab:alpha_Fe_runs}. 

In addition to the 6 massive quiescent galaxies at $3<z<5$, we also repeat the 5 fitting configurations on SNR per {\AA} $>500$ spatially stacked spectra of two local ($z<0.05$) quiescent galaxies from the SDSS MaNGA survey \citep{MANGA} as controls. Stacking is performed via an unweighted sum of the flux column for each spaxel, while the uncertainties are summed in quadrature. MaNGA-11835-9101 is chosen as a classical `red and dead' old elliptical galaxy, which has no measurable star formation at least within the past 3 Gyr. MaNGA-12514-3702 is selected as a post-starburst galaxy in \cite{Leung2024}, which underwent a period of increased star formation at $\approx1\,$Gyr in lookback time, and subsequently rapidly quenched. The post-starburst nature of this object provides a useful analogue to the early massive quiescent galaxies in our EXCELS sample.

\subsubsection{Adding variable $\alpha$-abundance stellar models to \textsc{Bagpipes}}
We adopt two SSP libraries that provide models with variable [$\alpha$/Fe] abundances. Based on the same stars as the empirical MILES spectral library, the sMILES SSP library provides semi-empirical model spectra over the rest-frame wavelength range from $3540-7410\,${\AA}, ranging from $[\alpha \mathrm{/Fe]}=-0.2$ to $+0.6\,$dex \citep{Knowles2023}. The sMILES models provide a good match to our fiducial \cite{Bruzual2003} models because of their matching wavelength range and spectral resolution ($\mathrm{FWHM}=2.5${\AA}). Additionally, both were constructed from the same suite of empirical stellar spectra. However, the sMILES library lacks the broader rest-frame UV and IR coverage of the \cite{Bruzual2003} models, which drastically limits the number of photometric points that can be included along with the EXCELS spectroscopy for each fit. This has the potential to introduce biases in the measurements of SFHs and other physical properties \citep[e.g.,][]{Pforr2012,Hunt2019}.

The second model library is version 2.3 of Binary Populations and Spectral Synthesis (BPASS), also ranging from $[\alpha\mathrm{/Fe]}=-0.2$ to $+0.6\,$dex \citep{Byrne2022,Byrne2025}. Due to their completely theoretical nature, these models span the full wavelength range probed by our \textit{JWST}+\textit{HST} photometry, as well as our EXCELS spectroscopy.

We re-scale both SSP libraries to the \cite{Asplund2009} solar abundances (see Appendix \ref{apx:abundance}), and implement them in \textsc{Bagpipes}, including the ability to vary [$\alpha$/Fe] as a free parameter. The prior used is uniform within $-0.2<[\alpha/\mathrm{Fe}]<0.6$. All of the other fitting parameters and priors are as described in Table \ref{tab:priors}.

By default, \textsc{Bagpipes} does not include \textsc{Cloudy} photoionisation models that vary in [$\alpha$/Fe] abundance. As we do not fit any galaxies that show noticeable current or recent ($\lesssim100$ Myr timescale) star formation according to their estimated SFHs from our fiducial fits (see Section \ref{sec:sfr_results} and Fig. \ref{fig:all_sfh}), we do not explore the production of $\alpha$-enhanced \textsc{Cloudy} models in this work (though this is something we plan to address in future work). Although we lack photoionisation models that vary in [$\alpha$/Fe] consistently with the SSP libraries, we opt to include photoionisation models computed from SSPs without [$\alpha$/Fe] variation. Fits based on sMILES SSPs use nebular models computed from \cite{Bruzual2003} models. Fits based on BPASS version 2.3 use nebular models computed from the earlier BPASS version 2.2.1. Their inclusion is only to disfavour solutions with very young stellar ages during SED fitting. Because we only fit the 6 galaxies with no current or recent star formation within 500 Myr, the impact of not using photoionisation models that vary in [$\alpha$/Fe] is negligible.

For both sets of models we fit our EXCELS spectra over the same $3540-7350$ {\AA} wavelength range and with the same emission-line masking as described in Section \ref{sec:fitting}. In addition, we also mask the calcium H and K lines at $\lambda_\mathrm{rest}\approx3950\,${\AA} for all 5 rounds of fitting with variable [$\alpha$/Fe]. These lines are useful tracers of stellar calcium abundance, which can be used to measure $\alpha$-enhancement. However, calcium in the ISM can also provide a substantial contribution to these lines \citep[e.g.,][]{Murga2015}, which has the potential to bias our measurements of [$\alpha$/Fe].

\subsubsection{Fitting with \textsc{Alf}-$\alpha$}
Another commonly used method to constrain $\alpha$-enhancement is by measuring individual elemental abundances using the Absorption Line Fitter code \citep[\texttt{alf,}][]{Conroy2012,Conroy2018}. However, the stellar models used in \texttt{alf} do not extend below an age of $1\,$Gyr, making it unsuitable for some of the galaxies in our sample. Instead, we adopt the more-recently developed \texttt{alf}-$\alpha$ code \citep{Beverage2025} based on \texttt{alf}, which also includes the sMILES SSP library as well as the standard \cite{Conroy2018} \texttt{alf} stellar models. Since historically most works in the literature have used \texttt{alf} with the \cite{Conroy2018} models to measure individual elemental abundances, for completeness we use \texttt{alf}-$\alpha$ to fit both the sMILES models and the \cite{Conroy2018} models to our emission-line-masked EXCELS spectra. The \texttt{alf}-$\alpha$ code also includes the ability to produce predictions from the \cite{Conroy2018} models for stellar populations younger than 1 Gyr through extrapolation. We opt to use this functionality, however it should be noted that this comes with the potential for significantly increased systematic uncertainties.

When fitting with the sMILES models, we pass \texttt{alf}-$\alpha$ the full MILES-wavelength-range EXCELS spectra after masking emission lines\footnote{We divide the wavelength range into four polynomial regions: $3540<\lambda_\mathrm{rest}<4600\;${\AA}, $4600<\lambda_\mathrm{rest}<5600\;${\AA}, $5600<\lambda_\mathrm{rest}<6600\;${\AA}, and $6600<\lambda_\mathrm{rest}<7650\;${\AA}. We set the polynomial degree of freedom as four for all regions.}, as described in Section \ref{sec:fitting} (the \texttt{alf}-$\alpha$ code does not include the ability to fit photometric data). With these models, \texttt{alf}-$\alpha$ assumes a relatively simple galaxy model with 10 free parameters. These are redshift, stellar age (assuming a single-burst SFH model), total stellar metallicity ($\log_{10}(Z_*/\mathrm{Z_\odot})$), [$\alpha$/Fe], stellar velocity dispersion, emission line strengths for the Balmer lines (assuming case-B recombination) and [O\,\textsc{iii}], velocity offset and dispersion of the emission lines, and a white noise scaling term (similar to $s$ in Table \ref{tab:priors}).

When fitting with the \cite{Conroy2018} models, we limit the fitted wavelength range to $4000-6400${\AA} and $8000-8800${\AA} in the rest frame, following \cite{Conroy2014}, which we call the ``\texttt{alf} range''\footnote{We divide the wavelength range into four polynomial regions: $4000<\lambda_\mathrm{rest}<4800\;${\AA}, $4800<\lambda_\mathrm{rest}<5600\;${\AA}, $5600<\lambda_\mathrm{rest}<6400\;${\AA}, and $8000<\lambda_\mathrm{rest}<8800\;${\AA}. We set the polynomial degree of freedom as four for all regions.}. The fit assumes a single burst SFH model and has 20 free parameters, including all parameters listed above in the \texttt{alf}-$\alpha$ fit using sMILES except for [$\alpha$/Fe], which is replaced with individual abundances for 10 elements, including Mg and Fe, and an additional varying effective temperature for the hot star component \citep{Conroy2018}. To provide a close comparison to this round of fitting, we also repeat our \textsc{Bagpipes} fits using the BPASS library, this time limiting our EXCELS spectra to the \texttt{alf} range. This also probes the effects of fitting different rest-frame wavelength ranges on the measured properties. 

\begin{table*}
    \centering
    \caption{The configurations of the 5 fitting runs we adopt to measure the stellar $\alpha$-abundances of our massive quiescent galaxies, detailed in Section \ref{sec:alpha_method}. For the \textsc{Bagpipes} run using the sMILES SSP library, due to the limited wavelength range of sMILES, only photometric points with filter profiles that lie fully within the MILES wavelength range could be included, which is typically 1-2 NIRCam band(s). The ``colour'' column refers to the colours shown in Figs \ref{fig:alpha_Fe_bars} and \ref{fig:alpha_Fe_delta}.}
    \begin{tabular}{llllll}
        \hline
        Fitting code         & SSP library & SFH model        & Wavelength range (rest frame)                            & Photometry             & Colour  \\ 
        \hline
        \texttt{alf}$\alpha$ & \cite{Conroy2018} (VCJ)         & single burst     & \texttt{alf} range: $4000-6400${\AA} \& $8000-8800${\AA} & None                   & black   \\
        \textsc{Bagpipes}    & BPASS v2.3  & double power law & \texttt{alf} range: $4000-6400${\AA} \& $8000-8800${\AA} & full \textit{JWST}+\textit{HST}        & blue    \\
        \textsc{Bagpipes}    & BPASS v2.3  & double power law & MILES range: $3540-7410${\AA}                        & full \textit{JWST}+\textit{HST}        & orange    \\
        \textsc{Bagpipes}    & sMILES      & double power law & MILES range: $3540-7410${\AA}                        & \textit{JWST}+\textit{HST} if in range & lime    \\
        \texttt{alf}$\alpha$ & sMILES      & single burst     & MILES range: $3540-7410${\AA}                        & None                   & magenta \\ \hline
    \end{tabular}
    \label{tab:alpha_Fe_runs}
\end{table*}

\subsection{Variable $\alpha$-abundance fitting results} \label{sec:alpha_results}

\begin{figure*}
    \centering
    \includegraphics[width=\textwidth]{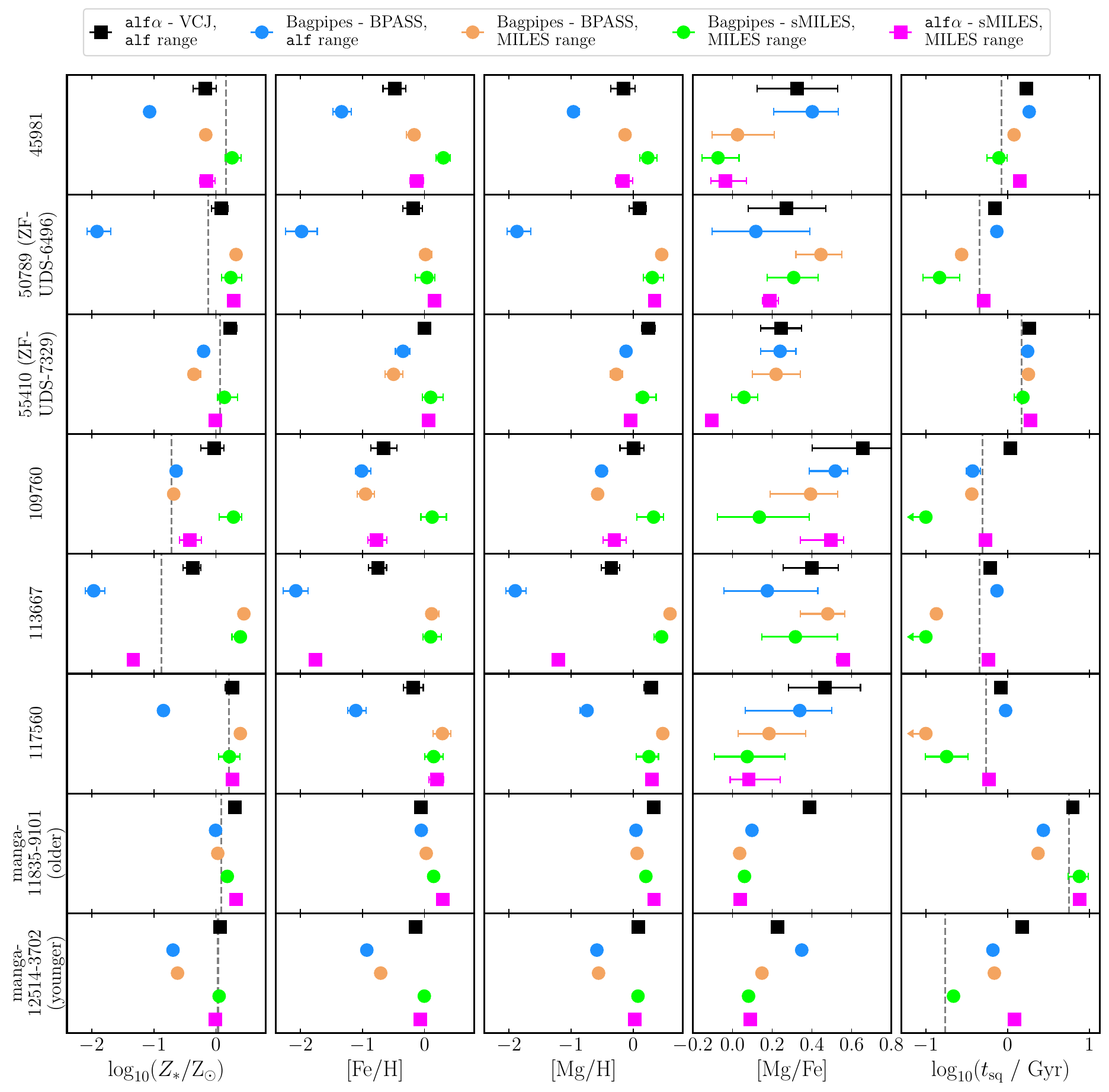}
    \caption{A comparison of the stellar properties estimated from our 5 different $\mathrm{[\alpha/Fe]}$-varying fitting configurations (see Section \ref{sec:alpha_method} and Table \ref{tab:alpha_Fe_runs}) for 6 massive quiescent galaxies from our sample and 2 MaNGA low-redshift control objects. Galaxies are ordered in rows, and stellar properties are ordered in columns. From left to right, columns show the total stellar metallicity, [Fe/H], [Mg/H], [Mg/Fe] and the lookback time to the point at which the galaxy quenched (defined as when $\mathrm{sSFR}<0.2/t_\mathrm{H}$ is satisfied for the first time). All abundances have been adjusted to the \protect\cite{Asplund2009} scale (see Appendix \ref{apx:abundance}). Within each panel, the marker and error bars denote the 16\textsuperscript{th}, 50\textsuperscript{th} and 84\textsuperscript{th} percentiles of the corresponding posterior distribution. The correspondence between colours and fitting runs is given in the figure legend and also in Table \ref{tab:alpha_Fe_runs}. Circles (squares) mark results from \texttt{alf}-$\alpha$ (\textsc{Bagpipes}) fits. Vertical dashed lines mark the posterior median estimate from the fiducial fits using \textsc{Bagpipes} and the \protect\cite{Bruzual2003} stellar models (not marked in the [Fe/H], [Mg/H] and [Fe/Mg] panels because these abundances were forced to follow the total stellar metallicity in the fiducial fits). The x-axis limits of the [Mg/Fe] column are set to $[-0.2,0.6]$, the limits of all \textsc{Bagpipes} fits. Instances where the posterior estimated SFH is not quenched at $z=z_\mathrm{obs}$ have $\log_{10}(t_\mathrm{sq}$/Gyr) marked as an upper limit at $-1$.}
    \label{fig:alpha_Fe_bars}
\end{figure*}

In Fig. \ref{fig:alpha_Fe_bars}, we compare the estimated values of $\log_{10}(Z_*/\mathrm{Z_\odot})$, [Fe/H], [Mg/H], [Mg/Fe] and time since quenching ($t_\mathrm{sq}$) from the 5 fitting configurations described in Section \ref{sec:alpha_method} and summarised in Table \ref{tab:alpha_Fe_runs}. We calculate the time since quenching as $t_\mathrm{sq} = t_\mathrm{H}(z_\mathrm{obs})\,-\,t_\mathrm{H}(z_\mathrm{quench})$, where $t_\mathrm{H}$ is the Hubble time, $z_\mathrm{obs}$ is the redshift at which the galaxy is observed, and $z_\mathrm{quench}$ is the redshift at which the galaxy first fell below sSFR = $0.2/t_\mathrm{H}(z)$. For fits using the sMILES library, we assume $\mathrm{[Mg/H]=[\alpha/H]}$ and $\mathrm{[Mg/Fe]=[\alpha/Fe]}$. For fits using the BPASS library, we perform linear interpolation based on an abundance table to obtain [Fe/H], [Mg/H] and [Mg/Fe] (see Appendix \ref{apx:abundance}). 
In all panels, the fitting configurations are colour-coded according to the legend at the top of the figure, with full details given in Table \ref{tab:alpha_Fe_runs}. Symbols mark the posterior median values, while the error bars mark the 16th$-$84th percentile ranges of the posterior distributions. Where applicable, the vertical dashed lines mark the posterior median estimates from the fiducial \textsc{Bagpipes} \cite{Bruzual2003} fits from Section \ref{sec:results} with no variable $\alpha$-abundance. All results are homogenised onto to the \cite{Asplund2009} solar abundance scale following the methods described in Appendix \ref{apx:abundance}.

\subsubsection{High-redshift sample} \label{sec:alpha_highz}
In the three left-most columns in Fig. \ref{fig:alpha_Fe_bars}, significant disagreements between the different fitting configurations can be seen. Estimated $\log_{10}(Z_*/\mathrm{Z_\odot})$, [Fe/H], and [Mg/H] values for the same $z>3$ galaxy can vary by as much as 2 dex between the configurations (e.g., 113667). It can be seen that fits based on different SSP libraries lead to significantly varying results. For example, results from both BPASS configurations (blue and orange) for 45981 and 55410 consistently show lower $\log_{10}(Z_*/\mathrm{Z_\odot})$, [Fe/H] and [Mg/H] estimates than the other configurations, as well as our fiducial \cite{Bruzual2003} estimate for $\log_{10}(Z_*/\mathrm{Z_\odot})$.
This is consistent with the results of \cite{Byrne2023}, who find significantly varying strengths for the same metal absorption lines at the same ages and metallicities in different model libraries, attributing this to the different libraries adopting varying ingredients and assumptions, such as stellar isochrones and spectral templates. This is also consistent with recent results at $z\simeq2$ from \cite{Jafariyazani2025}.

In addition, fits performed on different wavelength ranges of the observed spectra give substantially different results in some cases, even when fitting with the same model library. In Fig. \ref{fig:alpha_Fe_bars}, for 45981, 50789, 113667 and 117560, results from the \textsc{Bagpipes}-BPASS configuration fitting the \texttt{alf} wavelength range (blue) tend to show significantly lower $\log_{10}(Z_*/\mathrm{Z_\odot})$ than the \textsc{Bagpipes}-BPASS configuration fitting the MILES wavelength range (orange), along with lower [Fe/H] and [Mg/H], and slightly older ages (e.g., 50789). The direction of these differences is consistent with the well-documented age-metallicity degeneracy in spectral fitting, where an older, more metal-poor stellar population can produce a similar spectrum to a younger, more metal-rich stellar population \citep[e.g.,][]{Worthey1994,Conroy2013}. These differences are potentially attributable to the fact that the light from different stellar populations dominate at different wavelengths. Therefore, different portions of the galaxy spectrum are preferentially sensitive to different stellar populations, as well as different elemental abundances, due to the inclusion of different spectral features \citep{Conroy2010,Conroy2013,Baldwin2018}.

As can be seen in the second column from the right in Fig. \ref{fig:alpha_Fe_bars}, the [Mg/Fe] values we estimate for the early massive quiescent galaxies are typically highly uncertain for all fitting configurations. The best constraints are obtained for 55410, the galaxy with the highest SNR in our high-redshift sample. 
We also observe substantial scatter in estimated [Mg/Fe] across the different fitting configurations, 
however, due to the large uncertainties, the different [Mg/Fe] estimates typically agree within their $1\sigma$ uncertainties.

For 109760 (4th row from top in Fig. \ref{fig:alpha_Fe_bars}), good agreement is achieved among all fitting configurations except \textsc{Bagpipes}-sMILES (lime dot). \textsc{Bagpipes}-sMILES returns significantly higher $\log_{10}(Z_*/\mathrm{Z_\odot})$, [Fe/H] and [Mg/H], and crucially no measurable $t_\mathrm{sq}$ value. This is caused by an extremely poorly constrained SFH for the \textsc{Bagpipes}-sMILES fit, where the posterior median SFH rises gradually from $z>10$ to $z\sim5$, followed by a sharper decline to the time of observation. This broad, continuous SFH is in stark contrast to the results obtained in the fiducial fit and all other fitting configurations, suggesting that the limited wavelength baseline of the sMILES library, which greatly reduces the number of photometric bands that can be fitted, can seriously affect the accuracy and precision of quantities measured. In this case, assuming a simpler model such as a single burst in SFH can sufficiently reduce modelling complexity to return a more-accurate and better-constrained posterior estimate, as demonstrated by the \texttt{alf}-$\alpha$-sMILES configuration (pink square). 

From the left column of Fig. \ref{fig:alpha_Fe_bars}, it might appear that the \textsc{Bagpipes}-sMILES fitting configuration (lime) shifts the stellar metallicity of 109760 and 113667, 2/3 of the metal-poor galaxies from our fiducial fits discussed in Section \ref{sec:metallicity}, up to approximately solar values, in line with the other galaxies. However, as discussed above, the \textsc{Bagpipes}-sMILES results suffer from a reduced wavelength baseline, which leads to largely unconstrained SFHs. This affects both 109760 and 113667, hence their \textsc{Bagpipes}-sMILES metallicity estimates are not trustworthy.

Whilst it is clear that our spectral fitting results with variable [$\alpha$/Fe] abundances suffer from considerable systematics as discussed above, some sensible basic conclusions can be drawn. PRIMER-EXCELS-55410 (ZF-UDS-7329), an ultra-massive quiescent galaxy that assembled at $z\gtrsim10$, exhibits the smallest uncertainties and the best agreement between the fitting configurations in all measured quantities shown in Fig. \ref{fig:alpha_Fe_bars}. This galaxy is likely solar or slightly super-solar in [$\alpha$/Fe], which could indicate a moderately prolonged assembly time. This might be surprising when compared to this galaxy's measured SFH from our fiducial model, which suggests extremely early and relatively rapid assembly ($\tau_{10-90}\approx300\;$Myr, Fig. \ref{fig:all_sfh}). However, the fiducial $\tau_{10-90}$ estimate is highly uncertain ($\approx250\;$Myr). 

Despite their large uncertainties, the pair of ultra-massive quiescent galaxies at $z=4.62$, 109760 and 117560 are consistent with stronger $\alpha$-enhancement, with most fitting results for these objects favouring $\mathrm{[Mg/Fe]}>0.1$. This result is consistent with the extremely rapid formation and quenching required for the pair to be quiescent by only $\simeq1.3$ Gyr after the Big Bang. One slightly lower mass and lower redshift galaxy, 45981, appears to exhibit the weakest $\alpha$-enhancement within the sample, as would be expected. 

Lastly, it is worth noting that 113667 is one of the three objects discussed in Section \ref{sec:results} for which our fiducial \cite{Bruzual2003} fitting returns a very low stellar metallicity. This object displays the most severe disagreement between the different fitting runs in all measured quantities in Fig. \ref{fig:alpha_Fe_bars}, again suggesting that current stellar models are not well able to reproduce a subset of relatively young quiescent galaxies at $z>3$.

\subsubsection{Low-redshift MANGA control sample}
As shown in the two lower rows in Fig. \ref{fig:alpha_Fe_bars}, better agreement is achieved between the five fitting configurations when fitting the local controls. We also observe a significant reduction in the uncertainties in [Mg/Fe] compared to the high-redshift results, likely a result of the significantly higher SNR of the control spectra (SNR per {\AA} $>500$).

Particularly good agreement is reached between the fitting configurations for the older ``red and dead'' MaNGA-11835-9101, while some disagreement remains when fitting the younger post-starburst MaNGA-12514-3702 (BPASS provides slightly more metal-poor results than the other fits). This again indicates that the SSP libraries achieve better mutual agreement at older stellar ages \citep[e.g.,][]{Jones2025}.
Given the more-similar stellar age of MaNGA-12514-3702 to the younger high-redshift massive quiescent galaxies, the disagreements in the results from MaNGA-12514-3702 could be a reflection of the disagreements from our high-redshift sample, both caused by poorer consistency among SSP libraries at younger stellar ages.

\subsection{Can we reliably measure the stellar chemical abundance patterns of the earliest quiescent galaxies?}\label{sec:alpha_discussion}

In the previous section, we measured Fe and Mg abundances, total metallicities and $\alpha$-enhancements (using [Mg/Fe]) for 6 massive quiescent galaxies in our sample, using three different $\alpha$-varying SSP libraries, two different fitting codes and two different wavelength ranges. Despite only selecting the galaxies with moderately high SNR per {\AA} $>6$ and older stellar ages (> 500 Myr), we observe substantial disagreements between the results from the different fitting configurations in all estimated properties for most galaxies. Disagreement in total metallicity, [Fe/H] and [Mg/H] can reach $>1$ dex. Aside from the oldest and highest SNR galaxy, 55410, uncertainties in [Mg/Fe] from most fitting configurations are $>0.15$ dex, with the $1\sigma$ confidence interval in some cases spanning more than half of the prior space of $-0.2<\mathrm{[Mg/Fe]<0.6}$. Although disagreement significantly decreases when fitting a local galaxy with much higher SNR per {\AA} $>500$ and a much older stellar age, galaxies that recently quenched can still lead to considerable disagreement even in the very-high-SNR regime.

The sources of the disagreements and difficulties encountered when measuring chemical abundance patterns in high-redshift massive quiescent galaxies are fourfold:

\begin{enumerate}
    \item Differences between predictions from different SSP libraries
    \item Differences between results from different wavelength ranges
    \item Different SFH models assumed in different codes
    \item Low continuum SNR in \textit{JWST} high-redshift galaxy spectra
\end{enumerate}

Firstly, there remains poor agreement between the different $\alpha$-varying SSP libraries \citep{Byrne2023}. As seen in Fig. \ref{fig:alpha_Fe_bars}, $\log_{10}(Z_*/\mathrm{Z_\odot})$, [Fe/H] and [Mg/H] estimates obtained using BPASS are often much lower than the estimates from other fitting configurations and our fiducial \cite{Bruzual2003} results, particularly when only fitting the \texttt{alf} wavelength range (blue dots). From testing several spectral libraries on a large sample of star clusters, \cite{Asad2025} found that theoretical spectral libraries tend to return lower stellar age and metallicity estimates, particularly when fitting low-SNR spectra, consistent with our findings for the theoretical BPASS library. 
We note that our comparison between $\alpha$-varying SSP libraries would be made more complete, and this issue in particular could be more thoroughly addressed, by the inclusion of $\alpha$-MC \citep{Park2025}, which is a new, fully theoretical SSP library computed from consistently $\alpha$-enhanced isochrones and stellar spectral templates. However, the $\alpha$-MC library is not yet publicly available.

Differences between $\alpha$-abundance and Mg abundance could also contribute to the disagreement. Compared to other $\alpha$-elements (e.g., Ca), the build-up of Mg is more strongly dominated by core-collapse supernovae \citep{Kobayashi2020}. Therefore, when a galaxy has super-solar (sub-solar) [$\alpha$/Fe], the galaxy's [Mg/H] and [Mg/Fe] will be higher (lower) than its $\mathrm{[\alpha/H]}$ and $\mathrm{[\alpha/Fe]}$. This subtle difference could impact our results in Fig. \ref{fig:alpha_Fe_bars}, as we use slightly different approaches for evaluating the Mg abundances for different SSP libraries. We assume $\mathrm{[Mg/H]} = \mathrm{[\alpha/H]}$ and $\mathrm{[Mg/Fe]} = \mathrm{[\alpha/Fe]}$ for sMILES, convert $\alpha$-abundances to Mg abundances when using BPASS, and  report Mg abundances directly from the fits when using the \cite{Conroy2018} models. If we assume our sample generally have super-solar [$\alpha$/Fe], this slight difference in approach between the SSP templates could partially explain the higher [Mg/Fe] estimate for most of these objects returned by the \textsc{Bagpipes}-BPASS-MILES-range fits (orange dots), compared to the \textsc{Bagpipes}-sMILES fits (lime dots). It could also explain the higher [Mg/Fe] estimate for 55410 and 117560 from \texttt{alf}-$\alpha$ using the \cite{Conroy2018} models (black squares), compared to using sMILES (magenta squares).

Secondly, galaxy properties measured by fitting different rest-frame wavelength ranges can lead to diverging results. This divergence varies with stellar age. Therefore, it is important to measure chemical abundances using a wide wavelength baseline to avoid being sensitive to only spectral features of one element, and to avoid only fitting spectral regions that are sensitive to limited stellar spectral types and stages of stellar evolution.

Thirdly, fits performed using \texttt{alf}-$\alpha$ assume a different SFH model (single burst) compared to those using \textsc{Bagpipes} (double power-law). The increased flexibility from the extended SFH used in \textsc{Bagpipes} could lead to estimates that diverge from ones made assuming all stars formed in a coeval burst (such as our fits for 109760, as discussed in Section \ref{sec:alpha_results}). To test this, we repeated all \textsc{Bagpipes} $\alpha$-varying fits with single-burst SFH models. On average, when compared to their double-power-law counterparts, we found the measured $\log_{10}(Z_*/\mathrm{Z_\odot})$, [Mg/Fe] and $t_\mathrm{form}$ differ by $\approx0.3\,$dex, $\approx0.12\,$dex and $\approx0.15\,$Gyr, respectively. These magnitudes are comparable to some of the differences seen between \texttt{alf}-$\alpha$ and \textsc{Bagpipes} estimates in Fig. \ref{fig:alpha_Fe_bars} (e.g., comparing lime and magenta), thus the assumption of different SFH models plausibly contributes to the observed differences between results measured from the different fitting configurations described in Section \ref{sec:alpha_results}. While the single burst model used in \texttt{alf} and \texttt{alf}-$\alpha$ is a reasonable simplification to make at lower redshifts due to the old stellar age of the typical quiescent galaxy, one should be more careful in making the same assumption when fitting recently quenched galaxies at earlier times.

Lastly, the SNRs of the \textit{JWST} spectra obtained for these early massive quiescent galaxies remain too low, contributing to the large uncertainties and scatter in [Mg/Fe] in Fig. \ref{fig:alpha_Fe_bars}. 
Although we observe the best agreements between the fitting configurations in 55410, the level of precision achieved allows relatively little to be concluded beyond that it likely has a solar or slightly super-solar $\alpha$-abundance ratio. Therefore, we estimate the minimum observed frame SNR per {\AA} required for useful $\alpha$-abundance measurements to be $\gtrsim15$, slightly above that of our 55410 observation (which has SNR per {\AA} $\simeq12$).

In summary, when measuring detailed chemical abundances and $\alpha$-enhancement, it is desirable to always report results from more than one fitting configuration, ideally using various SSP libraries. A wide wavelength baseline should be used, and a minimal SNR per {\AA} $\gtrsim15$ in observed frame is needed. This corresponds to a SNR per resolution element of $\simeq100$ for medium-resolution ($R\simeq1000$) spectroscopy.

\begin{figure*}
    \centering
    \includegraphics[width=0.9\textwidth]{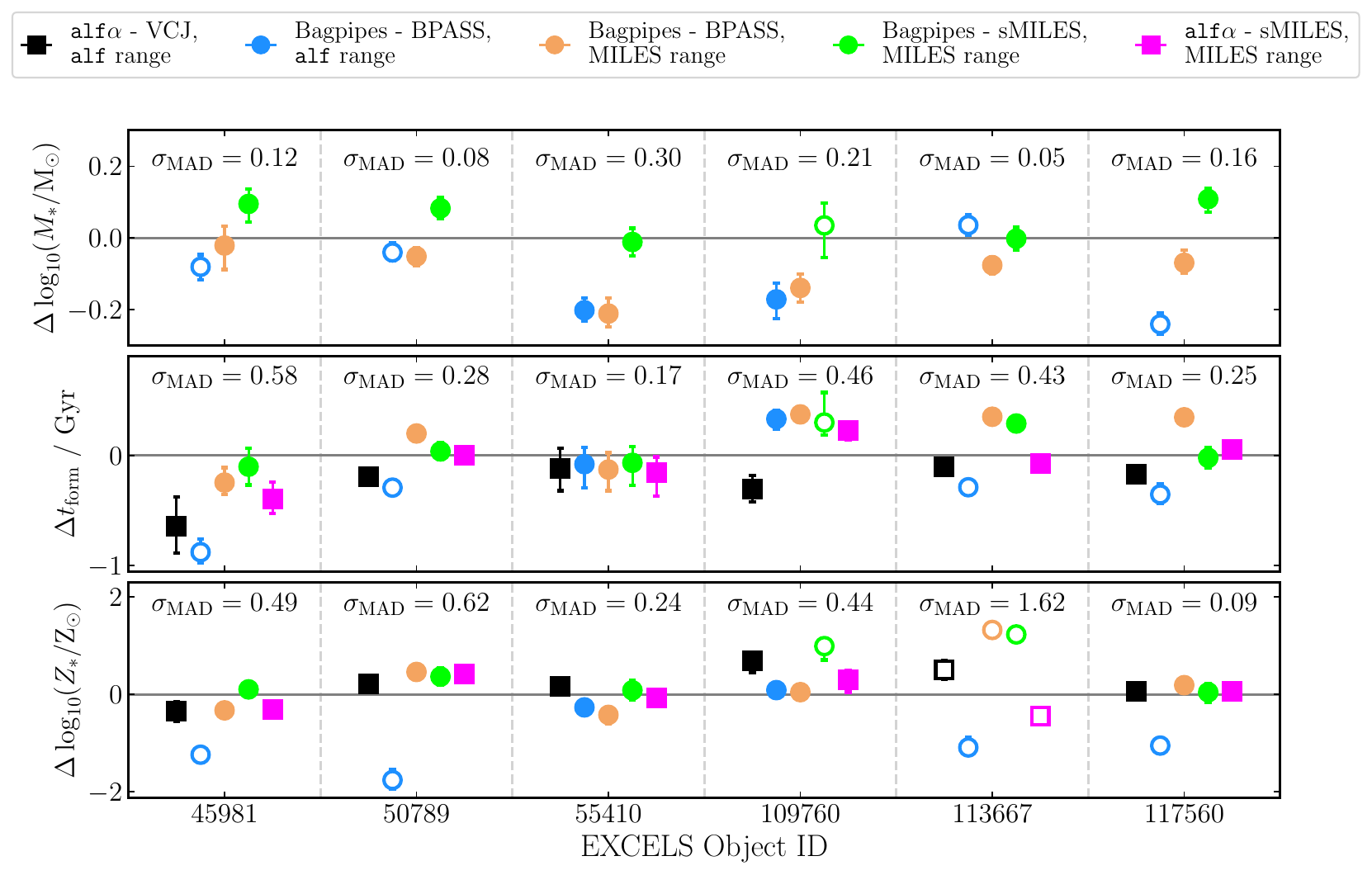}
    \caption{Differences in estimated galaxy physical properties between our fiducial \protect\cite{Bruzual2003} run and our 5 fitting configurations with variation in $\alpha$-abundances for the same 6 massive quiescent galaxies shown in Fig. \ref{fig:alpha_Fe_bars}. From top to bottom, the panels show the difference in stellar mass, the age of the Universe when 50 per cent of galaxy stellar mass had formed ($t_\mathrm{form}$), and stellar metallicity. Positive values indicate that the $\alpha$-varying estimate is higher than the fiducial result. Points with open symbols signify fits that were determined as obviously problematic in Section \ref{sec:alpha_highz}. Differences in stellar mass for \texttt{alf}-$\alpha$ fits are not shown because \texttt{alf}-$\alpha$ does not measure this parameter. In each panel, we also include the standard deviation (estimated via the median absolute deviation) of the differences in both properties for each galaxy, calculated across the 3 or 5 configurations. The colours again follow the scheme given in Fig. \ref{fig:alpha_Fe_bars} and Table \ref{tab:alpha_Fe_runs}.}
    \label{fig:alpha_Fe_delta}
\end{figure*}

\subsubsection{$\alpha$-abundance in PRIMER-UDS-55410 (ZF-UDS-7329)}

In Fig. \ref{fig:alpha_Fe_bars}, using \texttt{alf}-$\alpha$ with the \cite{Conroy2018} models, we estimate that the most massive galaxy in our sample, 55410 (ZF-UDS-7329), has $\mathrm{[Mg/Fe]}=0.24\pm0.1$. This is $\simeq0.2\;$dex lower than the measurement of $\mathrm{[Mg/Fe]}=0.42^{+0.19}_{-0.17}$ for this object using \texttt{alf} by \cite{Carnall2024}, but the two measurements are consistent within $1\sigma$ due to the large uncertainties. 

Although the two measurements show no significant tension, it will nonetheless be informative to investigate the source of the difference, which will help better our understanding of measurement systematics in the $\alpha$-abundance of the earliest quiescent galaxies. Four factors could contribute to this difference:
\begin{enumerate}
    \item Difference between the \texttt{alf} and \texttt{alf}-$\alpha$ codes
    \item Changes we made to the data reduction process downstream of the \textit{JWST} pipeline, namely the wavelength-varying 1D optimal extraction, joining of the three gratings and spectrophotometric calibration
    \item Changes we made to the data reduction process during running the \textit{JWST} pipeline, namely the addition of the \texttt{clean\_flicker\_noise} step and the manual masking of bad pixels
    \item Changes in the NIRSpec reduction pipeline and reference files since early 2024
\end{enumerate}

We fitted the exact version of the spectrum for 55410 used in \cite{Carnall2024} with \texttt{alf}-$\alpha$ and the \cite{Conroy2018} models, yielding $\mathrm{[Mg/Fe]}=0.35\pm0.1$, which is more consistent with the value given by \texttt{alf} in \cite{Carnall2024} than with the fiducial estimate we report based on the new spectrum. This suggests the switch from \texttt{alf} to \texttt{alf}-$\alpha$ only introduced a minor difference, thus rejecting \textit{reason 1} as a significant contributor to the difference.

To test \textit{reason 2}, we produce an alternative spectrum for 55410 using the same level 3 output 2D spectra produced in Section \ref{sec:pipeline}, but following the method of \cite{Carnall2024} when performing 1D optimal extraction, joining of the three gratings and spectrophotometric calibration. Fitting this new spectrum using \texttt{alf}-$\alpha$ and the \cite{Conroy2018} models yields $\mathrm{[Mg/Fe]}=0.17\pm0.1$, which is consistent with the result in Fig. \ref{fig:alpha_Fe_bars} using the same code and model configuration, but again different to the \cite{Carnall2024} result. This rejects \textit{reason 2} as the cause of the disagreement with \cite{Carnall2024}.

To test \textit{reason 3}, we repeat the reduction of 55410 following Section \ref{sec:pipeline}, but do not turn on the \texttt{clean\_flicker\_noise} step and do not perform manual masking of bad pixels. Comparing this spectrum to that used throughout this study, the difference remains $<1$ per cent at all wavelength bins not affected by any manual masking. For pixels affected by manual masks, differences at $\sim5$ per cent are typically found, but none of these wavelength bins are located at the key Mg and Fe absorption features found at $5100 < \lambda_\mathrm{rest} < 5400${\AA} for this galaxy. Therefore \textit{reason 3} is not the cause of the disagreement.

Finally, in Appendix \ref{apx:reduction_versions}, we show that changes made to the default NIRSpec reduction pipeline and calibration reference files since the version used in \cite{Carnall2024} led to a gradual shift in the slope of the spectrum (colour) and introduces finer changes in the reduced spectrum at the $5-10$ per cent levels. Some of these finer differences lie in wavelength regions that are sensitive to Mg and Fe abundances, which we demonstrate to have a significant impact on the estimated [Mg/Fe] value. We thus conclude that the disagreement is largely due to changes in the NIRSpec reduction pipeline and calibration reference files since early 2024.

\subsection{Impact on stellar mass, age and total metallicity estimates}\label{sec:alpha_conclusion}
We finally discuss the impact that fitting $\alpha$-enhancement as a free parameter can have on estimates for stellar mass, age and total metallicity, which could influence the conclusions we have drawn in Section \ref{sec:results}. As discussed in Section \ref{sec:metallicity}, \cite{Beverage2025} have reported that assuming scaled-solar abundances can produce up to $\simeq0.7$ dex biases in total stellar metallicities (median offset $\simeq0.4$ dex), compared to methods that measure individual chemical abundances. However, this bias is not observed when total metallicity from solar-scaled fits are compared to [Fe/H] instead, indicating the solar-scaled fits are more sensitive to Fe-peak abundance instead of $\alpha$-abundance. Here, we expand the investigation of this issue to our wider variety of code, model and wavelength-range configurations.

In Fig. \ref{fig:alpha_Fe_delta}, we plot the difference between the stellar mass, $t_\mathrm{form}$ and $\log_{10}(Z_*/\mathrm{Z_\odot})$ estimates returned by each $\alpha$-varying fitting configuration and the fiducial results from the BC03 fits. 
Several fitting results with outlying $\Delta$ values in one or more of the three quantities can be clearly seen, namely \textsc{Bagpipes}-BPASS-\texttt{alf} range for 45981, 50789, 113667 and 117560, and \textsc{Bagpipes}-sMILES for 109760. These fits have already been determined as obviously problematic above in Section \ref{sec:alpha_highz}, and are marked in Fig. \ref{fig:alpha_Fe_delta} with open symbols. Additionally, as noted in Section \ref{sec:results}, the metallicity of PRIMER-EXCELS-113667 prove to be challenging to measure accurately. Therefore, we also do not consider this galaxy's $\Delta\log_{10}(Z_*/\mathrm{Z_\odot})$ values.
Once the outliers are removed, Fig. \ref{fig:alpha_Fe_delta} shows that we do not observe a measurable global bias towards higher or lower stellar mass, $t_\mathrm{form}$ or $\log_{10}(Z_*/\mathrm{Z_\odot})$. Excluding the outliers, when alpha abundance is varied, offsets do not exceed 0.2 dex in stellar mass and 0.5 Gyr in $t_\mathrm{form}$.

We measure the typical $\Delta t_\mathrm{form}$ between $\alpha$-enhanced and non-$\alpha$-enhanced runs by measuring the standard deviation via the median absolute deviation across all configurations in all galaxies, obtaining $\sigma_\mathrm{MAD}=0.31\,$Gyr. This is considerably smaller than the range spanned by our sample (Fig. \ref{fig:simple_results}). Thus, the lack of $\alpha$-variation in our fiducial fits does not strongly affect our conclusions in Sections \ref{sec:results} and \ref{sec:downsizing} concerning downsizing. 

Similarly, our outlying results are removed from the bottom panel of Fig. \ref{fig:alpha_Fe_delta}, $\log_{10}(Z_*/\mathrm{Z_\odot})$ estimates from the fiducial fit have offsets not exceeding $0.7\,$dex, in good agreement with the maximum offsets observed by \cite{Beverage2025}. We measure $\sigma_\mathrm{MAD}=0.5\,$dex, slightly larger than the $\simeq0.3$ dex scatter reported by \cite{Beverage2025}. However, \cite{Beverage2025} reported that $\log_{10}(Z_*/\mathrm{Z_\odot})$ measured when varying $\alpha$-enhancement are offset by $\sim+0.4\,$ dex compared to non-$\alpha$-enhanced measurements. We do not observe such an offset in our results from Fig. \ref{fig:alpha_Fe_delta}. For the fits that assume solar abundance mixtures, we use \textsc{Bagpipes} along with the 2016 version of the \cite{Bruzual2003} models, while \cite{Beverage2025} used \textsc{Prospector} along with FSPS models \citep{Conroy2009,Leja2019a}. Since in Section \ref{sec:alpha_discussion} we have shown that using different SSP models can lead to considerably different estimates of stellar metallicity, it is possible that this difference in the level of offset is due to differences between the \cite{Bruzual2003} and FSPS models. This highlights that, whilst we are aiming to investigate the impact of fitting $\alpha$-enhancement as a free parameter, in practice it is challenging to disentangle this from the effects of switching SSP model libraries.

\subsection{Comparisons with other $z>3$ massive quiescent galaxy samples}
\cite{Hamadouche2026} measured the alpha abundance of their sample of 10 massive quiescent galaxies using \texttt{alf}-$\alpha$ and the sMILES models, finding the galaxies to be generally alpha-enhanced (median [$\alpha$/Fe] $\approx 0.2$), but with a large scatter. From this sample alone, the authors do not observe correlations between [$\alpha$/Fe] and either stellar mass or formation duration. However, when compared to galaxies at $1<z<3$ \citep{Slob2024,Beverage2025} and $z\sim0.7$ \citep{Beverage2021}, the authors observe a negative trend between [$\alpha$/Fe] and cosmic time, thus concluding that their sample of $z>3$ massive quiescent galaxies are generally more alpha-enhanced because of their earlier formation.

For the 6 EXCELS massive quiescent galaxies discussed in this section we find large scatter between [Mg/Fe] estimates, with a median\footnote{All 5 fitting configurations included to calculate the median.} [Mg/Fe] $\simeq0.24$, consistent with \cite{Hamadouche2026}. We do not find any statistically significant correlations between [Mg/Fe] and any of stellar mass, formation time and formation duration within this sample. Given the uncertain and model-dependent nature of the [Mg/Fe] estimates in these early massive quiescent galaxies as discussed above, we caution that any comparisons to lower redshift samples or correlations with other stellar properties might be subject to the effect of considerable systematics, especially when only fitting with a single SSP library.

\section{Conclusions}\label{sec:conclusion}
We have investigated the evolution and quenching of massive galaxies in the first 2 billion years of cosmic time with a sample of 14 massive quiescent galaxies at $3<z<5$ benefiting from extremely deep, medium-resolution ($R\simeq1000$) continuum spectroscopy at $\lambda=1-5\,\mu$m from the \textit{JWST} EXCELS survey. We perform a full, customised re-reduction of the EXCELS spectroscopy, introducing a novel wavelength-varying 1D optimal extraction method that combats the spectral ``wiggles'' caused by undersampling of the NIRSpec point spread function in the cross-dispersion (spatial) direction (see Section \ref{sec:wiggles}). This reduces the amplitude of the ``wiggles'' from $\simeq5$ per cent to $\simeq1$ per cent in the most problematic cases.

We analyse the spectroscopic data, along with the available multi-wavelength \textit{HST}+\textit{JWST} photometry, using Bayesian full spectra fitting, producing our main set of fiducial results with the \cite{Bruzual2003} stellar population models. We obtain good constraints for 12 galaxies with the other 2 objects having insufficient SNR. 

We find that massive quiescent galaxies at $3<z<5$ exhibit a tight negative correlation between their stellar mass and formation time (see Fig. \ref{fig:simple_results}), indicating that more massive galaxies assembled the bulk of their stellar mass earlier than less massive ones. Thus, the ``downsizing'' trend well known at lower redshift was already in place by $z\sim4$. We fit the slope of the stellar mass-formation time relation, finding that it is consistent with those derived from various samples of spectroscopically observed massive quiescent galaxies at $0<z<3$ from the literature (see Fig. \ref{fig:results}), at $\simeq2$ Gyr per decade in stellar mass (Equation \ref{eq:age_vs_mass}).
It is particularly noteworthy that we do not find any lower-mass relic galaxies in our sample: no objects with $\log_{10}(M_*/\mathrm{M_\odot})<10.8$ have formation times earlier than $\simeq1$ Gyr after the Big Bang. Therefore we conclude that $\log_{10}(M_*/\mathrm{M_\odot})\simeq10$ quiescent galaxies at higher redshift, such as the $z=7.3$ quiescent galaxy recently reported in \cite{Weibel2025}, will likely rejuvenate on timescales of a few hundred Myr.

The fitted SFHs suggest that most early massive quiescent galaxies experienced extremely rapid assembly of their stellar mass, forming 80 per cent of their stellar mass in only $\sim200\,$Myr. Their high peak SFRs ($>300\,\mathrm{M_\odot\,yr^{-1}}$) are comparable to the SFRs of the most extreme submillimetre galaxies at higher redshifts ($z\gtrsim5$).

The majority of our sample have relatively high stellar metallicities, comparable with results for literature massive quiescent galaxies from $0 < z < 3$ (see Figs \ref{fig:simple_results} and \ref{fig:results}), though we find no clear stellar mass-metallicity relation from our relatively small sample. Two objects however are fitted as much more metal poor, at $\log_{10}(Z_*/\mathrm{Z_\odot})\approx-0.7$. This has also been reported for some of the other highest redshift massive quiescent galaxies in the literature \citep{Carnall2023c,deGraaff2025,Wu2025,Weibel2025}, potentially indicating a new evolutionary pathway for massive, early galaxies that rapidly formed and quenched whilst maintaining substantially sub-solar metallicities. However, as no such objects are found at lower redshift, we caution that it is also highly plausible that these results are due to an inadequacy of current stellar models in the relatively little explored $\simeq500$ Myr age range that many $3 < z < 5$ massive quiescent galaxies inhabit. We test fitting these objects whilst imposing higher stellar metallicities, finding no significant impact on our derived stellar mass-formation time relationship.

Next, and partly motivated by these results, we investigate the detailed stellar chemical abundances of our sample, in particular their $\alpha$-enhancements, which we measure as [Mg/Fe] (our fiducial \citealt{Bruzual2003} fits assume scaled-solar abundances). For a sub-sample of the oldest 6 galaxies with higher SNR, we have tested 5 fitting configurations with varying combinations of fitting code, $\alpha$-abundance-varying SSP library, and fitted wavelength range (see Section \ref{sec:alpha_method} and Table \ref{tab:alpha_Fe_runs}). We find considerable disagreement between the total metallicities and abundance ratios measured by the 5 fitting configurations (see Section \ref{sec:alpha_results} and Fig. \ref{fig:alpha_Fe_bars}). These differences are contributed to by deviations between the predictions of different SSP libraries, by differing fitted wavelength ranges, different assumed SFH models, and the limited wavelength baseline of some SSP libraries, as well as showing some evidence of being exacerbated by younger galaxy stellar ages (see Section \ref{sec:alpha_discussion}).

We observe large uncertainties in most of our estimated [Mg/Fe] abundance ratios, suggesting that spectra with higher SNR than is typical for our sample are required for robust measurements of $\alpha$-enhancement. We suggest that future high-redshift studies should aim for observed-frame SNR per {\AA} $\gtrsim15$ over $5100<\lambda_\mathrm{rest}<6400\,${\AA}.
Despite these challenges, from comparing the $\alpha$-varying results with our fiducial scaled-solar abundance results using \cite{Bruzual2003}, we find that the assumption of the solar abundance mixture (scaled-solar abundances) likely introduces only fairly limited biases into our fiducial estimates of galaxy stellar mass, formation times ($t_\mathrm{form}$) and total stellar metallicity (see Section \ref{sec:alpha_conclusion} and Fig. \ref{fig:alpha_Fe_delta}).

Measuring detailed stellar chemical abundances for the earliest quiescent galaxies therefore remains highly challenging. The models that such measurements rely on (SSP libraries including variable non-solar abundance mixtures) are still in the early stages of development. We therefore suggest that future observational works aiming to measure detailed stellar elemental abundances should report estimates from more than one model library fitted to a wide wavelength baseline, thus mitigating (or at least exposing) potential biases caused by only using one fitting configuration. Considerably higher-SNR continuum spectra for the earliest quiescent galaxies at $3 < z < 5$ will also be needed to obtain precise measurements of their detailed chemical abundances.
Current medium-resolution data from surveys such as EXCELS are however sufficient to measure the stellar ages of such systems, and to provide a first indication of their total stellar metallicities, 
allowing us to begin moving towards understanding how stellar mass assembly and the quenching of star formation took place in the early Universe. 

\section*{Acknowledgements}

We thank Massissilia Hamadouche, \'Angel Chandro-G\'omez and Yingjie Peng for providing data. We thank Conor Byrne for providing BPASS abundance tables and helping with its conversions. HL thanks Anne Sansom and Elizabeth Stanway for discussions that helped with interpreting the $\alpha$-enhancement results. HL thanks Sirio Belli for useful discussions regarding results. HL, ACC, ET and SDS acknowledge support from a UKRI Frontier Research Grantee Grant (PI Carnall; grant reference EP/Y037065/1). FC, KZA-C, DS and TMS acknowledge support from
a UKRI Frontier Research Guarantee Grant (PI Cullen; grant reference EP/X021025/1).
VW acknowledges Science and Technologies Facilities Council (STFC) grants ST/V000861/1 and ST/Y00275X/1, and Leverhulme Research Fellowship RF-2024-589/4. OA acknowledges the support from STFC grant ST/X006581/1. JSD and DJM acknowledge the support of the Royal Society through the award of a Royal Society University Research Professorship to JSD.

This research is based in part on observations made with the NASA/ESA Hubble Space Telescope obtained from the Space Telescope Science Institute, which is operated by the Association of Universities for Research in Astronomy, Inc., under NASA contract NAS 5–26555. 
This work is based in part on observations made with the NASA/ESA/CSA James Webb Space Telescope. The data were obtained from the Mikulski Archive for Space Telescopes at the Space Telescope Science Institute, which is operated by the Association of Universities for Research in Astronomy, Inc., under NASA contract NAS 5-03127 for JWST. These observations are associated with programs \#1837 and \#3543. The authors acknowledge the PRIMER team for developing their observing program with a zero-exclusive-access period. Support for Program number JWST-GO-03543.014 was provided through a grant from the STScI under NASA contract NAS5-03127.

\textit{Software:} \textsc{Astropy} \citep{astropy}, \textsc{Bagpipes} \citep{Carnall2018,Carnall2019b}, \textsc{Celerite2} \citep{Foreman-Mickey2017,Foreman-Mickey2018}, \textsc{Marvin} \citep{marvin}, \textsc{Matplotlib} \citep{matplotlib}, \textsc{Nautilus} \citep{nautilus}, \textsc{Numba} \citep{numba}, \textsc{Numpy} \citep{numpy}, \textsc{pipes\_vis} \citep{pipes_vis}, \textsc{Scipy} \citep{scipy}, \textsc{Seaborn} \citep{seaborn}

For the purpose of open access, the author has applied a Creative Commons Attribution (CC BY) licence to any Author Accepted Manuscript version arising from this submission.


\section*{Data Availability}
All \textit{JWST} and \textit{HST} data used in this work are available via the Mikulski Archive for Space Telescopes: DOI 10.17909/8acc-5324 (\url{https://mast.stsci.edu}). The reduced, extracted, calibrated and joined 1D spectra for all 14 galaxies, along with a machine readable version of Table \ref{tab:main_results} are available at \url{https://github.com/HinLeung622/EXCELS_massive_quiescent}. Additional data products are available from the authors upon request.



\bibliographystyle{mnras}
\bibliography{biblist} 



\appendix

\section{Star-formation histories} \label{apx:SFH}
\begin{figure*}
    \centering
    \includegraphics[width=\textwidth]{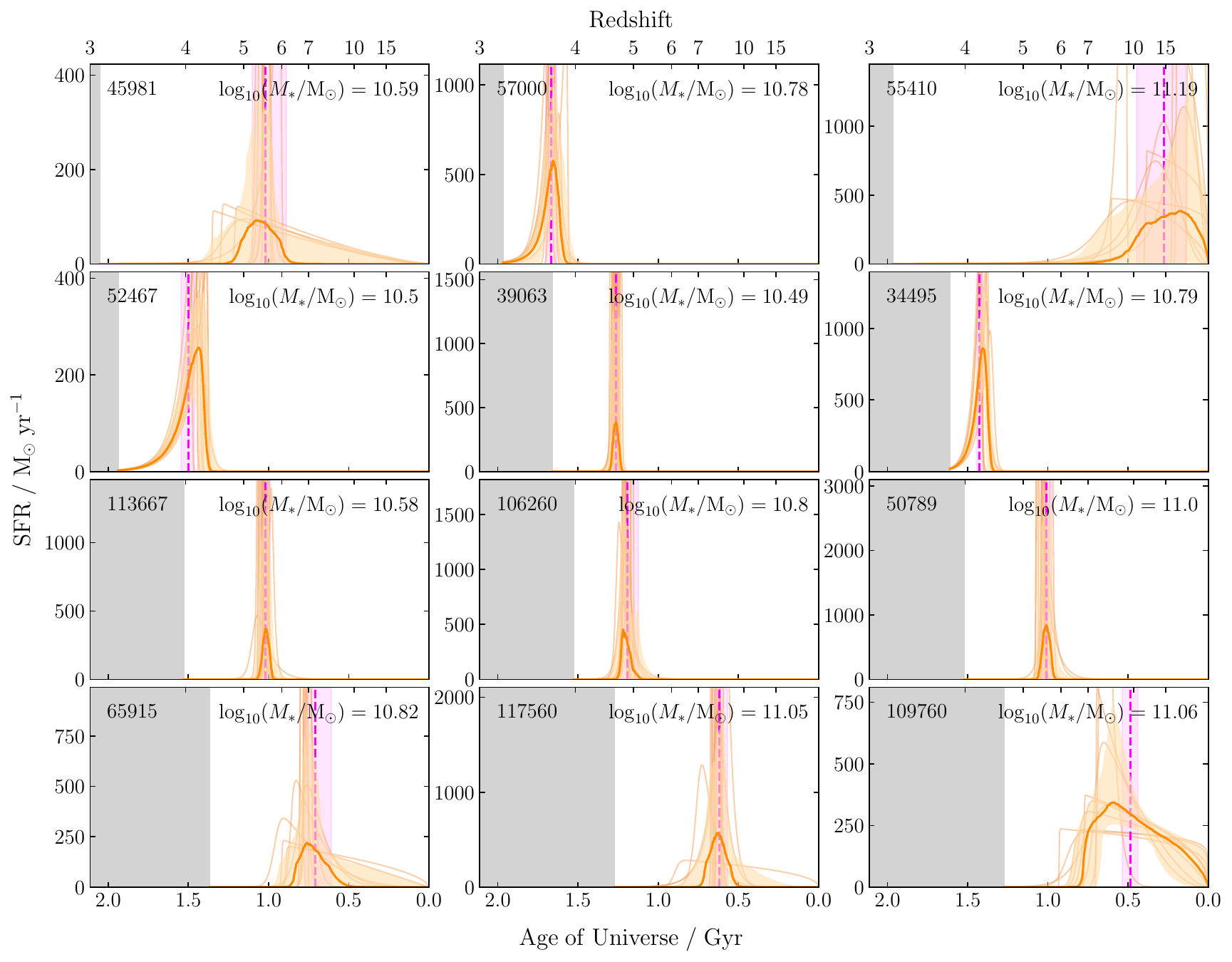}
    \caption{Fitted star-formation histories for 12 of our $3 < z < 5$ massive quiescent galaxies, obtained from \textsc{Bagpipes} full spectral fitting. The posterior median SFHs (thick orange lines) are plotted alongside their 16th$-$84th percentile confidence intervals (orange shaded regions). We also show the SFHs of ten random samples drawn from the posterior distributions with lighter orange lines. The vertical pink dashed lines mark the posterior median times of formation ($t_\mathrm{form}$), while the pink shaded regions mark their $1\sigma$ uncertainties. Cosmic time after each galaxy's observation is shaded grey.}
    \label{fig:all_sfh}
\end{figure*}

Fig. \ref{fig:all_sfh} shows the fitted SFHs of the 12 massive quiescent galaxies for which we obtain good constraints from the \textsc{Bagpipes} full-spectra-fitting approach described in \ref{sec:fitting}. The 2 galaxies with low SNR and missing coverage around $\lambda_\mathrm{rest}\sim4000\,${\AA} are not shown.

\section{Chemical abundance scale conversions} \label{apx:abundance}
Various SSP libraries often assume different solar metallicity scales and abundance mixtures. To compare results across these libraries properly, these values must first be converted into a common scale. For this work, we choose the solar metallicity scale and abundance mixture from \cite{Asplund2009}. For Fig. \ref{fig:alpha_Fe_bars} we require the abundance ratios [Fe/H], [Mg/H] and [Mg/Fe], while the \textsc{Bagpipes} fitting methodologies described in Section \ref{sec:alpha_method} only directly measure $\log_{10}(Z_*/\mathrm{Z_\odot})$ and $[\alpha\mathrm{/Fe}]$. Additional conversion is thus needed. Appendix \ref{sec:conversion_BPASS} details the conversion for BPASS v2.3, while Appendix \ref{sec:conversion_sMILES} details the conversion for sMILES. The \cite{Conroy2018} models are already on the \cite{Asplund2009} scale, and \texttt{alf-$\alpha$} directly provides all abundance ratios plotted in Fig. \ref{fig:alpha_Fe_bars}, so no conversions are required. 

In Table \ref{tab:solar_abundance}, we list the solar abundance values from two studies relevant in the rest of this section. For this section, we denote the expression $12+\log_{10}(N_\mathrm{X}/N_\mathrm{H})$ as $A_\mathrm{X}$,  where $N_\mathrm{X}$ and $N_\mathrm{H}$ are the number of atoms of the element in concern and Hydrogen per unit volume, respectively.

\begin{table}
    \centering
    \caption{Solar metal abundances from two commonly used solar chemical composition studies. The individual elemental abundances are given in the form $12+\log_{10}(N_\mathrm{X}/N_\mathrm{H})$ where X is the element. Here, we denote this form as $A_\mathrm{X}$.}
    \begin{tabular}{p{1.5cm}p{1.1cm}p{1.2cm}ll}
        \hline
        Paper & Proto-solar $Z$ & Photosphere $Z$ & $A_\mathrm{Fe,\odot}$ & $A_\mathrm{Mg,\odot}$ \\
        \hline
        \cite{Asplund2005} & 0.0130 & 0.0122 & $7.45\pm0.05$ & $7.53\pm0.09$ \\
        \cite{Asplund2009} & 0.0142 & 0.0134 & $7.50\pm0.04$ & $7.60\pm0.04$ \\
        \hline
    \end{tabular}
    \label{tab:solar_abundance}
\end{table}

\subsection{The BPASS library} \label{sec:conversion_BPASS}
BPASS v2.3 provides an abundance table in their public release\footnote{Found in \texttt{2025\_abundances.gz} in \url{https://warwick.ac.uk/fac/sci/physics/research/astro/research/catalogues/bpass/bpassv2p3/}.}. For each SSP, this table lists its metallicity (in mass fraction form), [$\alpha$/Fe], and the abundances of several key elements in the form $A_\mathrm{X}$. This includes values for H, Fe and Mg. [$\alpha$/Fe] is given calibrated to the \cite{Asplund2009} abundance scale.

Therefore, to perform the conversion from given $\log_{10}(Z_*/\mathrm{Z_\odot})$ and $[\alpha\mathrm{/Fe}]$ values to [Fe/H], [Mg/H] and [Mg/Fe], we first convert $\log_{10}(Z_*/\mathrm{Z_\odot})$ to metallicity in mass fraction:
\begin{equation}
    \log_{10}(Z_*) = \log_{10}(Z_*/\mathrm{Z_\odot}) + \log_{10}(\mathrm{Z_{\odot,A09}}) \; ,
\end{equation}
where $\mathrm{Z_{\odot,A09}}=0.0142$ is the proto-solar metallicity from \cite{Asplund2009}. Next, we perform 2D interpolations given the BPASS abundance table to obtain $A_\mathrm{Fe}$ and $A_\mathrm{Mg}$:
\begin{align}
    A_\mathrm{Fe} &= f_\mathrm{Fe}\Big(\log_{10}(Z_*), [\alpha/\mathrm{Fe}]\Big) \\
    A_\mathrm{Mg} &= f_\mathrm{Mg}\Big(\log_{10}(Z_*), [\alpha/\mathrm{Fe}]\Big) \;.
\end{align}
This interpolation is performed using the \texttt{scipy} \texttt{LinearNDInterpolator} routine. We then subtract the \cite{Asplund2009} solar abundance ratios in Table \ref{tab:solar_abundance} to get
\begin{align}
    [\mathrm{Fe/H}] &= A_\mathrm{Fe} - A_\mathrm{Fe,\odot,A09} \\
    &= A_\mathrm{Fe} - 7.50 \\
    [\mathrm{Mg/H}] &= A_\mathrm{Mg} - A_\mathrm{Mg,\odot,A09} \\
    &= A_\mathrm{Mg} - 7.60 \; .
\end{align}
Finally, the Mg to Fe abundance ratio can be directly calculated by
\begin{equation}
    [\mathrm{Mg/Fe}] = [\mathrm{Mg/H}] - [\mathrm{Fe/H}] \; .
\end{equation}

\subsection{The sMILES library} \label{sec:conversion_sMILES}
The chemical abundances of sMILES SSPs are labelled in terms of $\log_{10}(Z_*)$ and $[\alpha/\mathrm{Fe}]_\mathrm{A05}$, on the \cite{Asplund2005} scale. To perform our conversions, we make use of equation 2 in \cite{Knowles2023}, which relates [Fe/H], [$\alpha$/Fe] and total metallicity:
\begin{equation} \label{eq:knowles_eq2}
    [\mathrm{M/H}]_\mathrm{A05} = [\mathrm{Fe/H}]_\mathrm{A05} + a[\alpha/\mathrm{Fe}]_\mathrm{A05} + b[\alpha/\mathrm{Fe}]_\mathrm{A05}^2 \; ,
\end{equation}
where all abundances are on the \cite{Asplund2005} scale, and the total metallicity assumes the photospheric value $Z_\odot=0.0122$. \cite{Knowles2023} found the coefficients to be $a=0.66154\pm0.00128$ and $b=0.20465\pm0.00218$. Following \cite{Knowles2023}, $[\mathrm{M/H}]$ is defined as
\begin{equation}
    [\mathrm{M/H}] = \log_{10}(Z/X)_* - \log_{10}(Z/X)_\odot \; .
\end{equation}
This equation can be rearranged to give
\begin{equation} \label{eq:bulk_metallicity}
    [\mathrm{M/H}] = \log_{10}\Big(\frac{Z_*}{X_*} \cdot \frac{\mathrm{X_\odot}}{\mathrm{Z_\odot}} \Big) \; .
\end{equation}
We make the simplifying assumption that the variation in hydrogen mass fraction is negligible, such that $X_*\sim\mathrm{X_\odot}$, Equation \ref{eq:bulk_metallicity} therefore simplifies to
\begin{equation} 
    [\mathrm{M/H}] \approx \log_{10}(Z_*/\mathrm{Z_\odot}) \; .
\end{equation}

To use Equation \ref{eq:knowles_eq2}, all input values must first be converted to the appropriate \cite{Asplund2005} scale. For sMILES, we perform spectral fitting with [$\alpha$/Fe] kept in \cite{Asplund2005} scale, but total metallicity on the \cite{Asplund2009} scale (proto-solar). Therefore, we convert total metallicity to the \cite{Asplund2005} scale by:
\begin{equation} \label{eq:bulk_metallicity2}
    \log_{10}(Z_*/\mathrm{Z_\odot})_\mathrm{A05} = \log_{10}(Z_*/\mathrm{Z_\odot})_\mathrm{A09} + \log_{10}\bigg(\frac{\mathrm{Z_{\odot,A09}}}{\mathrm{Z_{\odot,A05}}}\bigg) \; .
\end{equation}
Then, we substitute Equation \ref{eq:bulk_metallicity2} into Equation \ref{eq:knowles_eq2} to calculate $[\mathrm{Fe/H}]_{\mathrm{A05}}$. 

\cite{Knowles2023} note that during the construction of sMILES, empirical [Mg/Fe] measurements of MILES stars are taken as a proxy for [$\alpha$/Fe]. Hence, for the purpose of the conversion, we assume $[\mathrm{Mg/Fe}] = [\alpha/\mathrm{Fe}]$. It follows that
\begin{equation}
    [\mathrm{Mg/H}]_\mathrm{A05} = [\alpha/\mathrm{Fe}]_\mathrm{A05} + [\mathrm{Fe/H}]_\mathrm{A05} \; .
\end{equation}
Finally, we convert from the \cite{Asplund2005} scale to the \cite{Asplund2009} scale by
\begin{align}
    [\mathrm{Fe/H}]_\mathrm{A09} &= [\mathrm{Fe/H}]_\mathrm{A05} + A_\mathrm{Fe,\odot,A05} - A_\mathrm{Fe,\odot,A09} \\
    &= [\mathrm{Fe/H}]_\mathrm{A05} + 7.45 - 7.50 \\
    [\mathrm{Mg/H}]_\mathrm{A09} &= [\mathrm{Mg/H}]_\mathrm{A05} + A_\mathrm{Mg,\odot,A05} - A_\mathrm{Mg,\odot,A09} \\
    &= [\mathrm{Mg/H}]_\mathrm{A05} + 7.53 - 7.60 \; .
\end{align}

\section{Differences in 1D spectra reduced from different \textit{JWST} pipeline versions}\label{apx:reduction_versions}
Motivated by the surprising difference between the [Mg/Fe] values for PRIMER-EXCELS-55410 measured by \cite{Carnall2024} and this work given the same observations, here we investigate if this difference could be caused by changes to the \textit{JWST} reduction pipeline. \cite{Carnall2024} used v1.12.5 of the pipeline and CRDS\_CTX=\texttt{jwst\_1183.pmap}, compared to our pipeline v1.19.1 and CRDS\_CTX=\texttt{jwst\_1413.pmap}. Therefore, we produce alternative versions of 1D optimally extracted spectra from the G235M grating for this galaxy from all major pipeline versions between v1.12.5 and v1.19.1. The reduction through each version is accompanied by the corresponding CRDS context version following table 1 of the \textit{JWST} Operations Pipeline Build Information webpage\footnote{\url{https://jwst-docs.stsci.edu/jwst-science-calibration-pipeline/jwst-operations-pipeline-build-information}} (see Table \ref{tab:reduction_versions}). In all reductions we also follow the default configurations in all three levels of the pipeline, meaning we do not include the \texttt{clean\_flicker\_noise} step and do not perform manual masking of bad pixels between running the level 2 and 3 pipelines. All 1D extraction were performed using the conventional optimal extraction method (wavelength-invariant, matching \citealt{Carnall2024}).

\begin{table}
    \centering
    \caption{\textit{JWST} pipeline versions and CRDS context files used to create alternative reduced 1D spectra of PRIMER-EXCELS-55410.}
    \begin{tabular}{ccc}
    \hline
         Pipeline version & CRDS\_CTX & Remarks\\
         \hline
         1.19.1 & 1413.pmap & Matches this study\\
         1.18.0 & 1364.pmap & - \\
         1.17.1 & 1321.pmap & - \\
         1.16.1 & 1303.pmap & - \\
         1.15.1 & 1293.pmap & - \\
         1.14.1 & 1240.pmap & - \\
         1.13.4 & 1210.pmap & - \\
         1.12.5 & 1183.pmap & Matches \cite{Carnall2024} \\
         \hline
    \end{tabular}
    \label{tab:reduction_versions}
\end{table}

\begin{figure*}
    \centering
    \includegraphics[width=\textwidth]{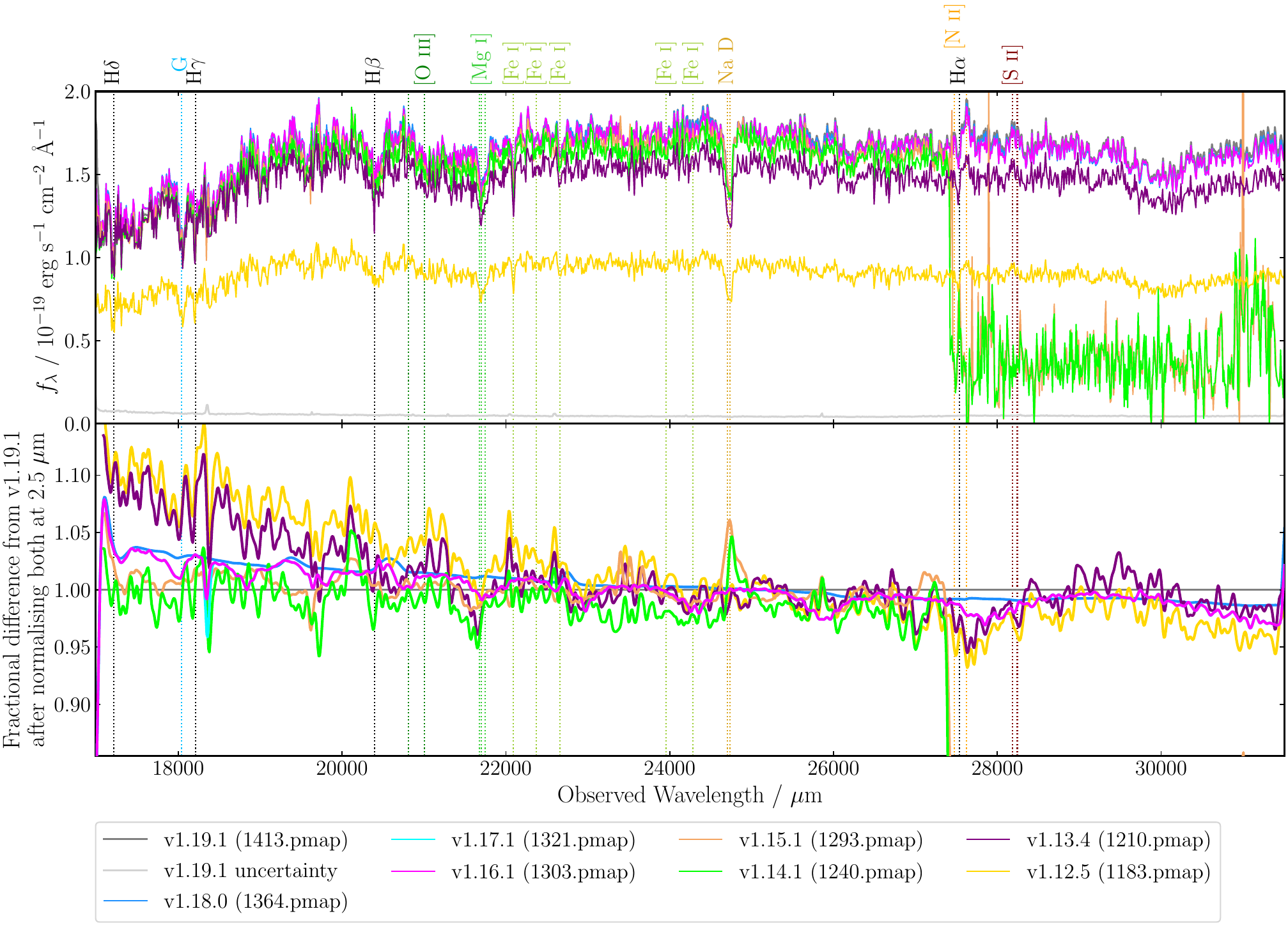}
    \caption{Comparing G235M 1D spectra produced by major \textit{JWST} NIRSpec reduction pipeline versions based on the same level 1 ``uncal'' input from the same observation of the same galaxy, PRIMER-EXCELS-55410. All reductions used their corresponding default settings. \textbf{Top}: The reduced 1D spectra, with colours following the legend at the top of the figure. The grey curve at the bottom of the panel marks the uncertainty from the v1.19.1 reduction. \textbf{Bottom}: All spectra in the top panel normalised by the flux at $\lambda_\mathrm{obs}=2.5\;\mu\mathrm{m}$ divided by the similarly normalised spectrum from v1.19.1. All curves are smoothed using a Gaussian blur with $\sigma=2$ wavelength bins. In both panels, we use vertical dotted lines to mark the expected location of key absorption and emission features for this galaxy's redshift ($z=3.195$).}
    \label{fig:reduction_compare}
\end{figure*}

In Fig. \ref{fig:reduction_compare}, we compare the 1D spectra produced by the various reduction pipeline versions. The top panel compares their fluxes. The middle panel shows the differences in the shapes of the spectra produced by the different pipeline versions. To produce the lines shown in this panel, we first normalise all spectra in the top panel by dividing each by their flux at $\lambda_\mathrm{obs}=2.5\,\mu\mathrm{m}$. Next, we take the ratio between each normalised spectrum and the v1.19.1 normalised spectrum. Finally, a Gaussian smoothing with $\sigma=2$ wavelength bins is applied before plotting.

The top panel of Fig. \ref{fig:reduction_compare} shows that there is a surprisingly large difference between 1D spectra reduced by the different NIRSpec pipeline versions since \cite{Carnall2024}, despite all starting from the same level 1 ``uncal'' products. The spectrum from v1.12.5 (yellow) is significantly dimmer than the spectra from all other versions for most of the wavelength range. This is likely due to a bug in the \textit{pathloss} step in the level 2 pipeline in v1.12.5 that prevented pathloss correction from being applied to MSA slitlets that are not in the conventional 1 by 1 or 1 by 3 configurations. This issue was fixed in v1.13.0 (pull request \#8106 on the \textit{JWST} pipeline github).

Versions 1.14.1 (lime) and 1.15.1 (orange) return spectra with significant missing flux at $\lambda_\mathrm{obs}>2.75\;\mu\mathrm{m}$. This is caused by the \texttt{outlier\_detection} step in the level 3 pipeline incorrectly labelling many of the pixels in the 2D spectra as outliers at these wavelengths. Thus, only a small fraction of pixels in the 2D spectra across all exposures at this wavelength range were combined into the final 2D spectrum, which leads to the missing flux. This issue appears to be solved in v1.16.1.

Beyond the obvious disagreements in normalisation between the spectra in the top panel of Fig. \ref{fig:reduction_compare}, the bottom panel shows that considerable disagreement in the spectral shape of the reduced 1D spectrum exists between the versions. The spectra from v1.12.5 (yellow) and v1.13.4 (purple) are brighter than that from v1.19.1 used in this work by up to 10 per cent in the blue end (after normalising both at $\lambda_\mathrm{obs}=2.5\;\mu$m). Even as recently as v1.18.0 (blue) the spectrum is 5 per cent brighter in the same wavelength range. Many finer differences between the spectra from different pipeline versions can also be seen, particularly when comparing the earlier versions to v1.19.1. These differences can have important implications for some measurements. For example, the significantly shallower Na D absorption line in the spectra reduced by v1.14.1 (lime) and v1.15.1 (orange) could heavily impact the measured neutral gas content and its possible outflow properties. The fractional curves from v1.17.1 (cyan) and v1.16.1 (magenta) show periodic patterns with periods akin to the ``wiggles'' discussed in Section \ref{sec:wiggles}. These patterns are no longer observed in the fractional curve from v1.18.0 (blue). This is caused by changes in v1.18.0 to the \texttt{barshadow} step in the level 2 pipeline (pull requests \#9085 and \#9326 on the \textit{JWST} pipeline github), which is dependent on the the shutter trace's projection on the detector shifting from one row to the next in the unrectified frame. One of the depressions of this periodic pattern coincides with the Mg \textsc{i} triplet at $\lambda_\mathrm{rest}\approx5175\;${\AA}, which boosts the absorption strength of these lines for spectra reduced with v1.17.1 or before.

\begin{figure}
    \centering
    \includegraphics[width=\columnwidth]{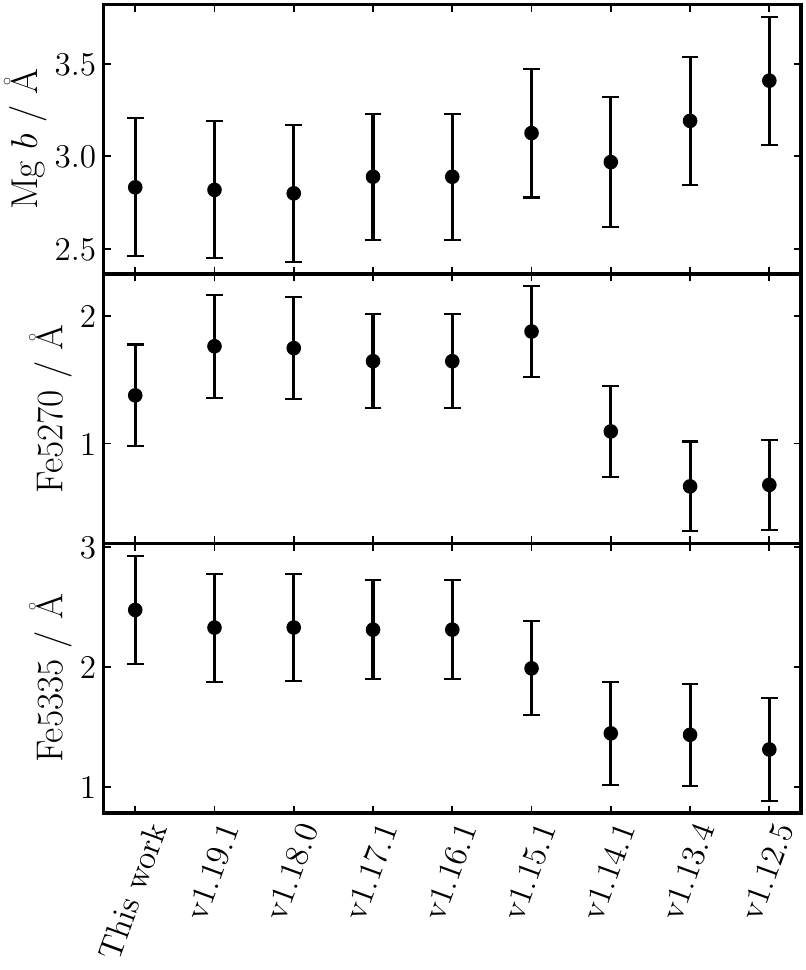}
    \caption{A comparison of the equivalent widths of three Mg or Fe sensitive Lick indices measured from spectra produced by various versions of the \textit{JWST} NIRSpec reduction pipeline for PRIMER-EXCELS-55410, alongside the spectrum produced by our custom reduction used in this work.}
    \label{fig:lick_indices}
\end{figure}

To investigate the impact of changes in the reduction pipeline on our [Mg/Fe] measurement of this galaxy, we measure the Lick indices sensitive to Mg and Fe abundances from all spectra shown in Fig. \ref{fig:reduction_compare}, along with the spectrum produced by our custom reduction used in this work. The measured indices are Mg \textit{b}, Fe5270 and Fe5335, following the bandpass definitions in \cite{Worthey1994}, and displayed in Figure \ref{fig:lick_indices}. Since v1.12.5, the changes in the reduction pipeline have led to shifts in the measured Lick index values at $\sim1\sigma$ levels. Mg b equivalent width decreased from v1.12.5 to v1.19.1, while Fe5270 and Fe5335 equivalent widths increased. This would result in lower Mg and higher Fe abundance estimates from spectra reduced with more recent pipeline versions, in agreement with the difference in [Mg/Fe] values measured using full spectral fitting between this work and \cite{Carnall2024} for this galaxy.

\bsp	
\label{lastpage}
\end{document}